\documentclass{aa}  

\usepackage{graphicx}
\usepackage{txfonts}
\begin{document}

   \title{Contribution of observed multi frequency spectrum of Alfv\'en waves to coronal heating}

   \author{P. Pagano
          \inst{1}
          \and
          I. De Moortel
          \inst{1}
          }

   \institute{School of Mathematics and Statistics, University of St Andrews, North Haugh, St Andrews, Fife, Scotland KY16 9SS, UK\\
      \email{pp25@st-andrews.ac.uk}
             }

  \titlerunning{Contribution of Alfv\'en waves to coronal heating}
  \authorrunning{Pagano \& De Moortel}

   \date{ }

 
  \abstract
   {Whilst there are observational indications that transverse MHD waves carry enough energy to maintain the thermal structure of the solar corona, it is not clear whether such energy can be efficiently and effectively converted into heating. Phase-mixing of Alfv\'en waves is considered a candidate mechanism,
   as it can develop transverse gradient where magnetic energy can be converted into thermal energy.
However, phase-mixing is a process that crucially depends on the amplitude and period of the transverse oscillations, and only recently have we obtained a complete measurement of the power spectrum for transverse oscillations in the corona \citep{Morton2016}.}
   {We aim to investigate the heating generated by phase-mixing of transverse oscillations triggered by buffeting of a coronal loop that follows from the observed coronal power spectrum as well as the impact of these persistent oscillations on the structure of coronal loops.}
   {We consider a 3D MHD model of an active region coronal loop and we perturb its footpoints with a 2D horizontal driver
   that represents a random buffeting motion of the loop footpoints.
   Our driver is composed of 1000 pulses superimposed to generate the observed power spectrum.}
   {We find that the heating supply from the observed power spectrum in the solar corona through phase-mixing is not sufficient to maintain the million degree active region solar corona.
   	We also find that the development of Kelvin-Helmholtz instabilities could be a common phenomenon in coronal loops
    that could affect their apparent life time.}
   {This study concludes that is unlikely that phase-mixing of Alfv\'en waves resulting from an observed power spectrum of transverse coronal loop oscillations can heat the active region solar corona. However, transverse waves could play an important role in the development of small scale structures.}
   

   \maketitle
%

\section{Introduction}
\label{introduction}

To explain the thermal structure of the solar atmosphere remains one of the main challenges for solar physicists. Although recent observations and theories have contributed to shedding light on the mechanisms behind the million Kelvin solar corona, a definitive explanation is currently out of reach
\citep[e.g.][]{ParnellDeMoortel2012,DeMoortelBrowning2015}.
The solar corona is a highly structured and dynamic environment where magnetic structures
\citep[such as coronal loops, e.g.][]{Reale2010}
are formed and dissipated throughout solar rotations and this complex evolution generates MHD waves in the solar corona in various way \citep[e.g.][]{Nakariakov1999,Tomczyk2007,DePontieu2007,McIntosh2011,Antolin2018}.
Some of these waves are observed to carry enough energy to significantly contribute to the energy budget of the solar corona, but it is not clear how the wave energy can be converted into thermal energy and other energy conversion mechanisms remain plausible,
such as the nanoflares model \citep[e.g.][]{Parker1988,Klimchuk2015}, for instance, in a scenario of the braiding of magnetic field lines \citep[e.g.][]{WilmotSmith2015}.
To make progress on the solution of the coronal heating puzzle, it is key to understand to what extent MHD waves contribute to the heating in order to estimate whether other mechanisms are essential or the combined effect of multiple mechanisms is needed to explain coronal heating.

It is now widely accepted that transverse waves are ubiquitous in the solar corona and that they carry a significant amount of energy \citep[e.g.][]{McIntosh2011,ParnellDeMoortel2012}.
More recently, \citet{Srivastava2017} have specifically looked at high-frequency propagating transverse waves and claim that enough energy to heat the corona and to sustain the solar wind is transferred to the solar corona.

However, while MHD waves can carry enough energy it has not been explained yet how this energy can actually be dissipated on the appropriate time and length scales \citep[see e.g.][]{Arregui2015}.
The phase mixing of Alfv\'en waves is a potential candidate to explain the dissipation of waves, as it naturally leads to small-scale structures (large gradients) in an inhomogeneous solar corona perturbed at the foot points.
\citet{Pascoe2010} and subsequent papers \citep{Pascoe2011,Pascoe2012} successfully demonstrated that the solar corona allows for the concentration of propagating wave energy at the boundary layers of dense structures and in follow up work, \citet{PaganoDeMoortel2017} assessed the physical range where phase-mixing can become relevant for coronal heating.
At the same time, other wave related mechanisms are being investigated, such as the generation of turbulence \citep{vanBallegooijen2011}
 or Kelvin-Helmholtz instabilities \citep[KHI, ][]{BrowningPriest1984, Terradas2008, Antolin2015, Howson2017, Pagano2018}.
All these mechanisms can be responsible for a portion of the energy input in the solar corona, where MHD waves are generated either from perturbations leaking from the photosphere and chromosphere \citep{Cally2003,Jess2009} or in\-situ when density flows interact \citep{Antolin2018}.
The results of these studies are not conclusive yet, as some models do not justify the
thermal energy deposition through damping of MHD waves
\citep{PaganoDeMoortel2017, Pagano2018}
and other models instead seem to have identified a plausible mechanism for that,
as \citet{LopezAristeFacchin2018} explained with an analytical model of how 3 minutes oscillations can lead to higher frequency modes in the corona 
and subsequently to the dissipation of enough energy to balance the radiative losses.

Past theoretical studies addressed how strongly the thermal energy deposition depends on the amplitude of the waves and on their period, where the temperature increase is larger with higher amplitude and short-period waves develop phase-mixing quicker \citep{HeyvaertsPriest1983, Hood1997}.
To this end, it is key to assess the heating effectiveness of this mechanisms not only from a theoretical point of view,
but also testing whether the heating following from the damping of the kind of oscillations present in the solar corona can
actually counteract the radiative losses of the coronal plasma.
\citet{Morton2016} measured the power spectrum of transverse oscillations in different regions of the solar corona using
the Coronal Multi-channel Polarimeter \citep[CoMP, ][]{Tomczyk2008}
and further empowered by the technique introduced by \citet{Weberg2018},
where an automated algorithm that identifies transverse oscillations and extract their wave properties is introduced.
They find that the power spectrum of the corona can be described with the superposition of three different components: a power law that describes long-periods oscillations (more than $\sim500$ s), a plateau near the 5 minutes oscillations, and a different power law for the short-period oscillations.
While it is unclear how this spectrum is generated, it is reasonable to consider this as a steady state power spectrum with which the solar corona oscillates.
If the conversion of wave energy into heating is sufficient to maintain the million degree solar corona, such energy is continuously extracted from 
this steady state spectrum whilst continuously being fed by wave energy, probably mostly from the chromosphere.

Therefore, to test this hypothesis, we model the effect of the observed spectrum on an idealised magnetic flux model to study the heating resulting from the dynamics triggered by the persistent buffeting of the loop footpoint
and how the loop structure is affected in this scenario.
We first explain how the observed spectrum in active regions can be generated from the interference of several pulses and what the consequences are in terms of the characteristics of the motion flows at the base of the solar corona.
Then a series of MHD numerical experiments where a coronal loop is modelled as a magnetised cylinder and one of the footpoint is displaced are used
to study the propagation of MHD waves along the structure.
We first study this dynamics with some selected single pulses, focusing on the development of electric currents and we then
use the full set of pulses, i.e. the complete spectrum to study the response of a coronal loop.
Finally we focus on the energy deposition and we also discuss how this affects the structures of coronal loops, especially with regards to KHI.

In Sect.\ref{multifdriver} we explain how the observed power spectrum is modelled,
in Sect.\ref{mhdevolution} we present some preliminary MHD simulations that are crucial to understand the results of our numerical experiments,
in Sect.\ref{solarsimulation} we present and analyse our MHD numerical experiments of a magnetised cylinder, and we discuss our results in Sect.\ref{discussions}.

\section{Multi frequency driver}
\label{multifdriver}
\begin{figure}
\centering
\includegraphics[scale=0.28]{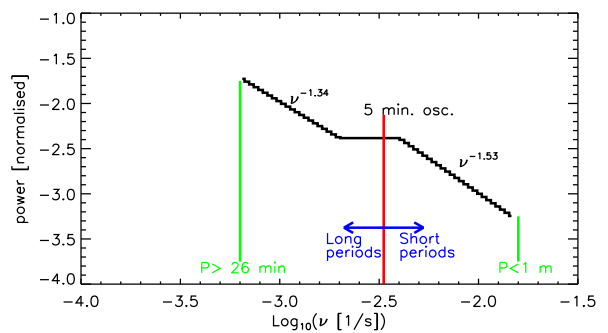}
\caption{Power spectrum of the transverse oscillations in the solar corona that we construct from the parameters for an active region in \citet{Morton2016}.}
\label{observedmodel}
\end{figure}

In order to study the energy deposition and effect of the propagation of MHD waves on a coronal loop, we use the result of \citet{Morton2016} to devise a model for the oscillatory behaviour of the loop.
\citet{Morton2016} derive the velocity power spectrum from transverse displacements
for different near limb regions of the solar corona
by measuring the Doppler velocity from the $10747$ $\AA$ Fe XIII line with CoMP.
While this research shows that the velocity power spectrum is different for different regions of the solar corona,
the key features are common across the solar corona.
The power spectrum is generally decreasing with frequency following a power law.
The power law index is different for frequencies below or higher than $\sim0.004$ Hz which corresponds
to the $\sim 5$ minutes period oscillations.
Around this frequency the velocity power spectrum does not 
behave according to a power law, but it shows an approximately flat power spectrum.
In our work, we adapt the power law derived by \citet{Morton2016} to study how the random horizontal buffeting motions of the loop footpoints induce propagating transverse waves along the loop and how much heating follows from this process.
Therefore, we first construct a set of pulses that mimic the measured power spectrum
and we subsequently generate a random motion-like movement from this set of pulses.
In particular, we use the parameters from \citet{Morton2016} for active regions to reproduce
our velocity spectrum shown in Fig.\ref{observedmodel} with an histogram with bins $\Delta Log_{10}(\nu)=0.023$ wide.

In our description we restrict our investigation to periods between
$\nu\sim 6.4 \times 10^{-4}$ Hz ($26$ minutes period) and
$\nu\sim 1.6 \times 10^{-2}$ Hz ($1$ minutes period),
and we use a description of the power law in segments:
\begin{equation}
W\left(\nu \right)=10^{\alpha}\nu^{\phi}
\label{modelspectrum}
\end{equation}
where $W\left(\nu \right)$ is the power spectrum, $\nu$ is the frequency and
the function parameters $\alpha$ and $\phi$
are given in Tab.\ref{tabalphaphi} for different $\nu$ ranges.
The resulting function is continuous at the frequencies where the regime changes.
\begin{table}
\caption{Power spectrum parameters}             
\label{tabalphaphi}      
\centering                          
\begin{tabular}{c c c}        
\hline\hline                 
$Log_{10}\left(\nu \right)$ range & $\alpha$ & $\phi$  \\    
\hline                        
   $-3.2 < Log_{10}\left(\nu \right) \le -2.7 $ & $-6.00$  & $-1.34$ \\  
   $-2.7 < Log_{10}\left(\nu \right) \le -2.4 $ & $-2.38$  & $0.00$  \\  
   $-2.4 < Log_{10}\left(\nu \right) \le -1.8 $ & $-6.05$  & $-1.53$ \\  
   
\hline                                   
\end{tabular}
\end{table}

\begin{figure}
\centering
\includegraphics[scale=0.28]{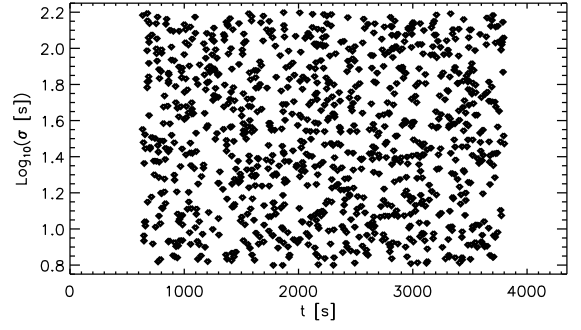}
\caption{Distribution of the 1000 pulses in the $t_0$ and $Log_{10}\left(\sigma [s]\right)$ space.}
\label{singlepulset0sig}
\end{figure}

We generate a power spectrum from the interference of a number of single pulses
over a certain time span, each of different duration, amplitude and occurrence time.
Each pulse is described by the following expressions where
$v_i\left(t\right)$ is the time derivative of $s_i\left(t\right)$
\begin{equation}
s_i\left(t\right)= -u_i\frac{\sigma_i}{\exp\left(-\frac{1}{2}\right)} \left[\exp\left[-\frac{(t-_{0i}-\epsilon\sigma_i)^2} {(2*\sigma_i^2)}\right]-\\
                            \exp\left[-\frac{(t-_{0i}+\epsilon\sigma_i)^2} {(2*\sigma_i^2)}\right]\right]
\label{displacement}
\end{equation}
\begin{equation}
u_i\left(t\right)= +\frac{u_i}{\sigma_i \exp\left(-\frac{1}{2}\right)} \left[\left(t-t_{0i}-\epsilon\sigma_i\right) 
                  \exp\left[-\frac{(t-t_{0i}-\epsilon\sigma_i)^2}{(2*\sigma_i^2)}\right]- \\
                                                \left(t-t_{0i}+\epsilon\sigma_i\right)
                  \exp\left[-\frac{(t-t_{0i}+\epsilon\sigma_i)^2}{(2*\sigma_i^2)}\right]\right]
\label{velocitydisplacement}
\end{equation}
where $u_i$ is the maximum velocity of the pulse,
$t_{0i}$ is the time when the pulse is centred,
$\sigma_i$ governs the time extent of the motion,
and we use $\epsilon=2$ to separate in time the two displacement peaks.
\begin{figure}
\centering
\includegraphics[scale=0.32]{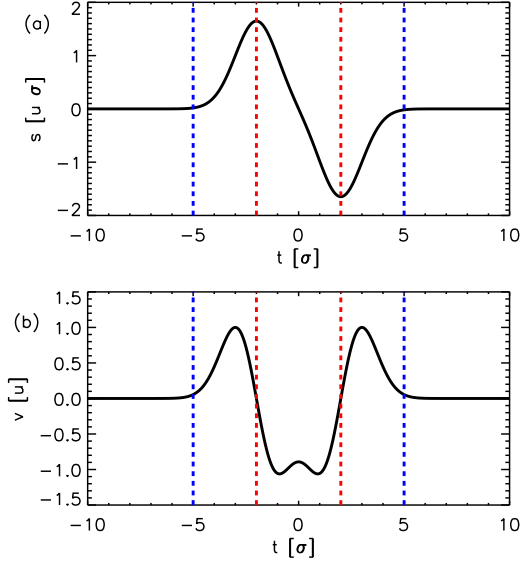}
\caption{(a) Displacement and (b) velocity profile of a single pulse oscillation (Eq.\ref{displacement}
and Eq.\ref{velocitydisplacement}).}
\label{singlepulse}
\end{figure}
Fig.\ref{singlepulse} shows the spatial displacement and the associated velocity for one pulse.
It should be noted that the velocity (Fig.\ref{singlepulse}b) peaks at $t=t_{0i}\pm(\epsilon\sigma+\sigma)$
and we associate a period $P_i=10\sigma_i$ to each pulse and a frequency $\nu_i=1/P_i$.
The parameters $t_{0i}$ and $\sigma_i$ are randomly chosen for $1000$ pulses
using a uniform distribution for $Log_{10}(\sigma)$ and
for $t_{0i}$ within a given time window to
ensure each pulse is fully contained
within the time frame of the simulation.
The resulting uniform distribution is shown in Fig.\ref{singlepulset0sig}.
Finally, in order to combine these pulses in a suitable way to reproduce the spectrum in Eq.\ref{modelspectrum},
we need to assign each pulse an appropriate amplitude.
The energy $E_i$ associated with a pulse is proportional to:
\begin{equation}
E_i\propto
\int^{+\infty}_{-\infty} u_i^2 \frac{t^2}{\sigma_i^2}e^{-\frac{t^2}{\sigma_i^2}} \, dt =
u_i^2\frac{\sqrt{\pi}}{2}\sigma
\label{powerpulse}
\end{equation}
and thus the power, $W_i$ (defined as the energy of the pulse divided by the period of the pulse) is $W_i\propto u_i^2$ and it is not dependent on $\sigma_i$.
Thus, we bin the pulses in a histogram
with $\Delta Log_{10} \nu=0.22$ as in Fig.\ref{observedmodel}
and we assign the amplitude $u_i$ to each pulse to match the intended power in each histogram bin.
Finally, we rescale the velocity amplitude $u_i$
of all pulses in order to have a maximum velocity of $10$ Km/s
for the combined driver,
which is an order of magnitude higher
than typical velocities in the solar corona due to transverse displacements \citep{Threlfall2013}.

\begin{figure}
\centering
\includegraphics[scale=0.28]{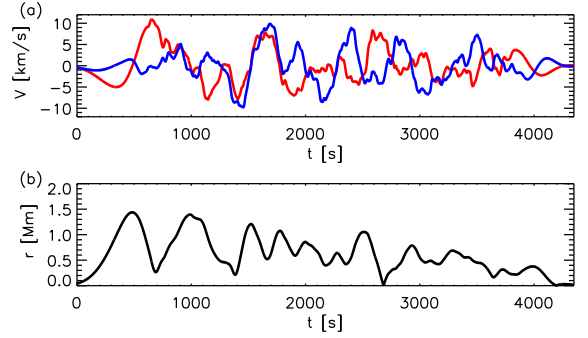}
\caption{(a) profile of the resulting $v_x(t)$ and $v_y(t)$ from the interference of 1000 pulses.
(b) resulting radial distance from the initial position as a function of time.}
\label{dispvtime}
\end{figure}

In order to obtain a horizontal driver that models
the random buffeting footpoint motion of a coronal loop footpoint, we associate a random angle $\theta_i$ to 
each pulse and thus we decompose the velocity as
$v_{xi}=u_i \cos\left(\theta_i \right)$ and
$v_{yi}=u_i \sin\left(\theta_i \right)$.
Fig.\ref{dispvtime} shows the resulting combined driver where we find that
the loop footpoint starts and returns to the rest position and
the x-velocity and y-velocity profiles consist of high-frequency
motions enveloped by higher velocity low frequency ones.

\begin{figure}
\centering
\includegraphics[scale=0.28]{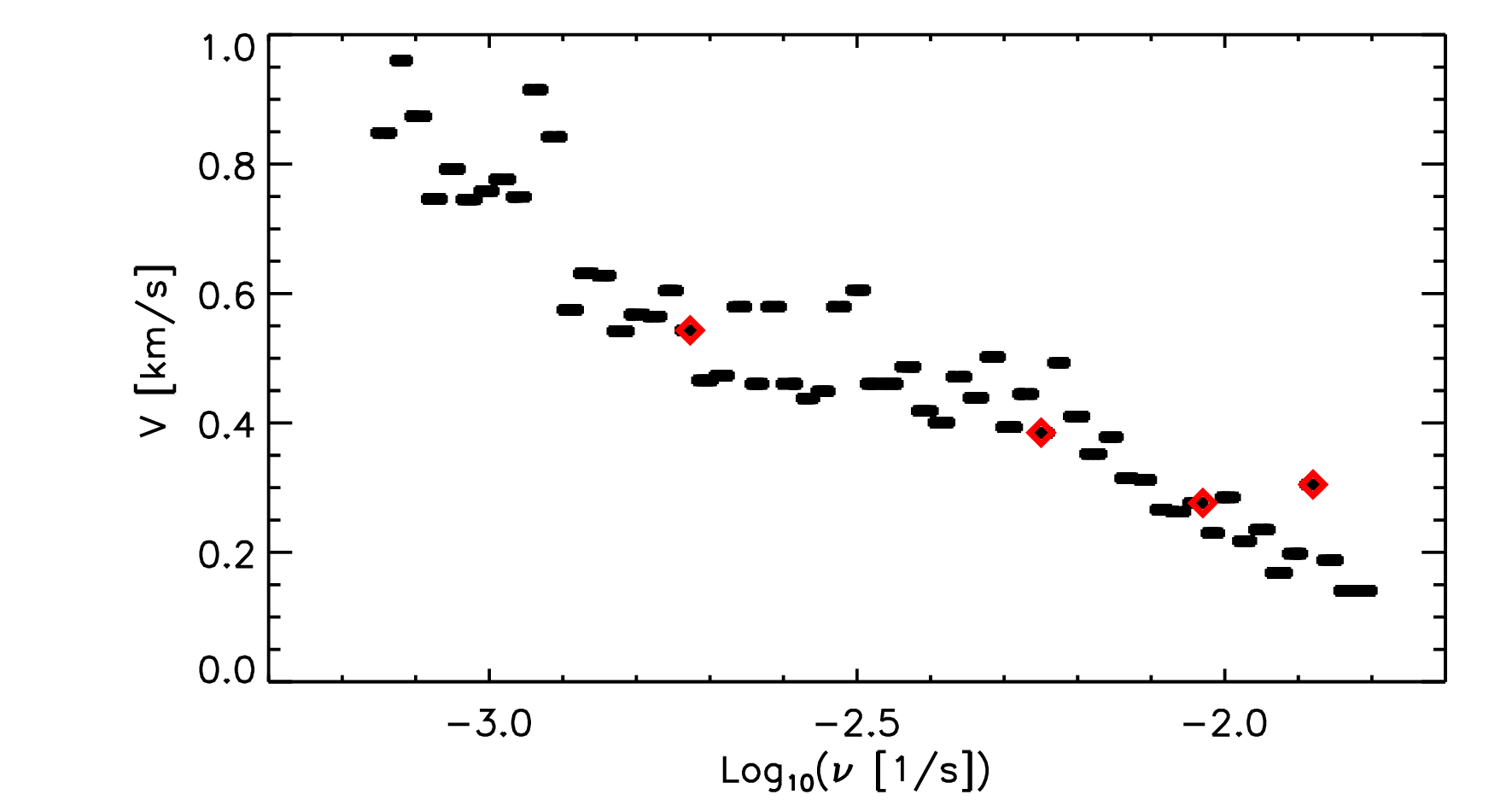}
\caption{Distribution of the pulses in the $u$,  $Log_{10}\left(\nu [1/s] \right)$ space.
The red points are the pulses we use for our preliminary simulations in Sect.\ref{mhdevolution}.}
\label{vmodfuncsig}
\end{figure}

Similarly, Fig.\ref{vmodfuncsig} shows the distribution of $u_i$
as a function of $Log_{10}\left(\nu_i \right)$ that follows from this power distribution, where we see that the amplitude of low frequency oscillations is about one
order of magnitude greater than the one of high-frequency oscillations.
The chosen modelling parameters are clearly a balance between a number of modelling requirements and observational constraints and are intended to be representative of active region loops.

\section{Monochromatic ideal single pulse simulations}
\label{mhdevolution}

In order to investigate the effect of the power spectrum constructed in Sec.\ref{multifdriver} on a coronal loop structure we devise an MHD model
perturbing a magnetised cylinder footpoint with the driver resulting from our modelling.  
Magnetised cylinders are wave guides for transverse MHD waves
and when footpoints are horizontally displaced, kink waves travel along the 
cylinder. If a continuous boundary (shell) region is present, mode-coupling with the azimuthal modes leading to an effective 
energy transfer from one wave mode to the other
\citep[e.g. ][]{RudermanRoberts2002,Pascoe2010}.
These Alfv\'en waves travel in a non-uniform medium
in this boundary layer and hence will phase-mix \citep[e.g.][]{HeyvaertsPriest1983,Pascoe2010,PaganoDeMoortel2017}.
A direct consequence of the phase-mixing is that the 
magnetic field is no longer constant along the direction 
perpendicular to the direction of propagation of the waves,
and thus electric currents are generated.
The generation of currents is key for the possible subsequent thermal energy deposition.

\begin{figure}
\centering
\includegraphics[scale=0.32]{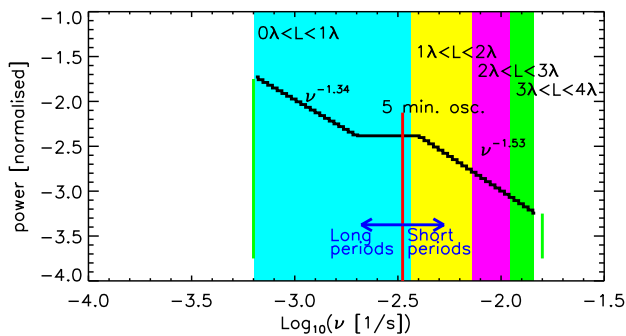} 
\caption{Power spectrum of the transverse oscillations in the solar corona that we construct from the parameters for an active region in \citet{Morton2016}. The different colours 
identify how many wavelengths of an Alfv\'en wave
travelling at $0.75$ Mm/s
fit in a loop that is $L=200$ Mm long.}
\label{observedmodelmaps}
\end{figure}

The more developed the phase-mixing is, the stronger the electric currents grow. For a flux tube of a given length, Alfv\'en waves with shorter wavelengths $\lambda=\nu V_A$ will lead to more developed phase-mixing. In this study, we consider a magnetised cylinder where $L=200$ Mm and Fig.\ref{observedmodelmaps} shows how many wavelengths can fit in a loop
for different frequencies when we
assume an Alfv\'en speed of $V_A=0.75$ Mm/s.
We find that the long-period power law and the 5-minute oscillations plateau 
correspond to waves whose wavelength is comparable to the full extent of the cylinder.
Only shorter period waves are able to oscillate two times or more.
On the other hand, the modulus of these currents is proportional to the modulus of the velocity and thus longer periods
from the velocity power spectrum described in Sec.\ref{multifdriver}
will produce stronger electric currents, just because their oscillation amplitude is larger.
In order to understand the interplay between these effects and how the resulting
electric currents are distributed along the magnetised cylinder
we run four ideal MHD simulations
where we pick four different pulses to drive the footpoint.
These pulses have different amplitudes and periods: $P=533$ s, $P=178$ s, $P=107$ s and $P=76$ s,
triggering Alfv\'en waves with wavelengths (where $L=200$ Mm is the length of the loop): $\lambda=2\,L$, $\lambda=2/3\,L$, $\lambda=2/5\,L$ and $\lambda=2/7\,L$
(red symbols in Fig.\ref{vmodfuncsig}).

\subsection{Numerical setup}
\label{simulationsetup}
To model our footpoint driven coronal loop, we set up a numerical experiment where a magnetised cylinder is composed of a dense interior region and 
a boundary shell across which the density decrease and the Alfv\'en speed increases. We then perturb the footpoint(s) of this cylinder to study the propagation of MHD waves.

\begin{figure}
\centering
\includegraphics[scale=0.5,clip,viewport=80 68 370 645]{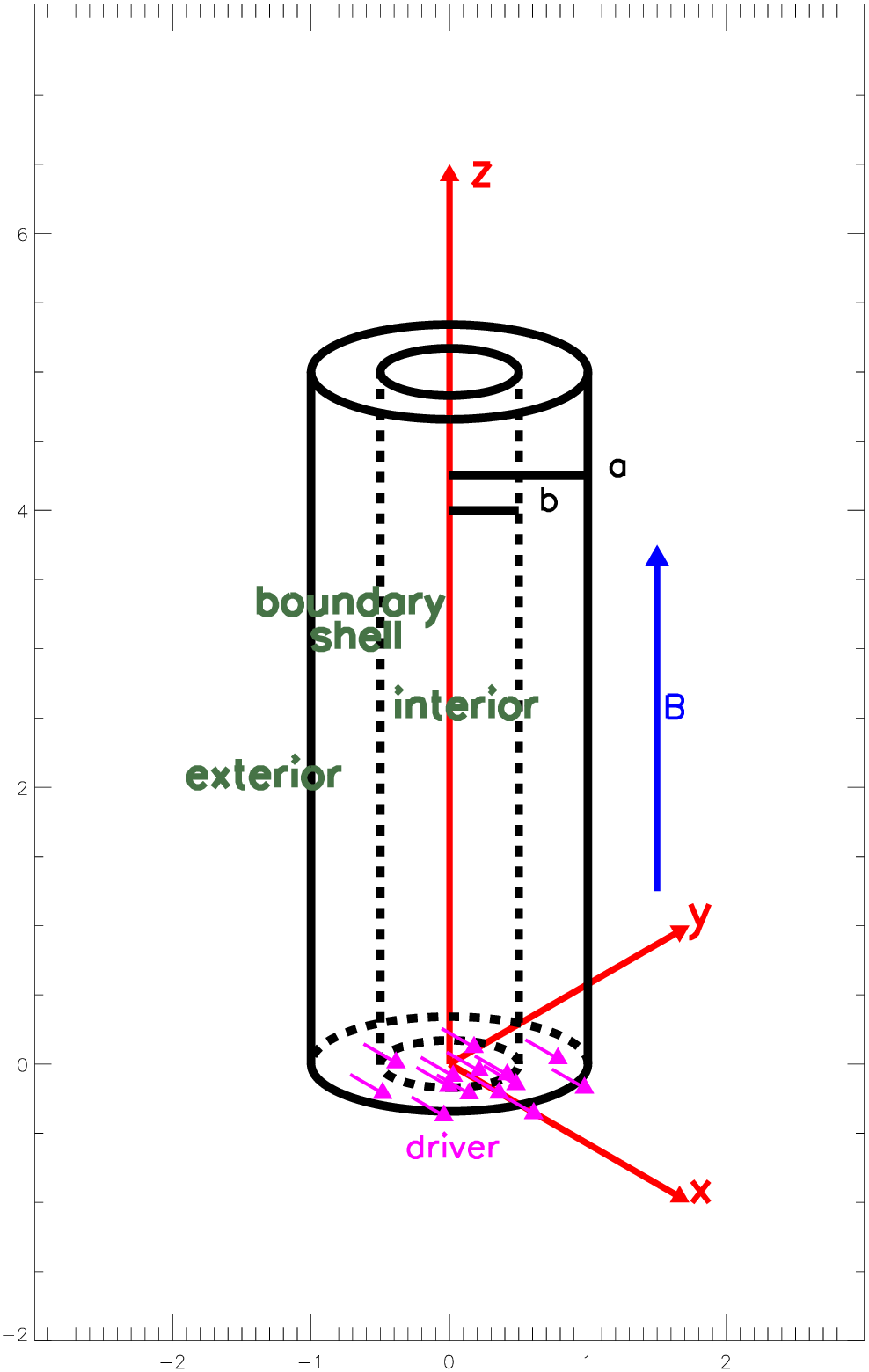}
\caption{Sketch to illustrate the geometry of our system and the Cartesian axes.}
\label{sketch}
\end{figure}
Fig.\ref{sketch} illustrates the model which is based on the model presented 
by \citet{Pascoe2010,Pascoe2011,Pascoe2012,PaganoDeMoortel2017}.
The system is built in a Cartesian reference frame 
with $z$ being the direction along the cylinder axis.
The origin of the axes is placed at the centre of the footpoint of the cylinder.
The cylinder has radius $a$,
the interior region has radius $b$.
The region between radii $b$ and $a$ is the boundary shell.
In our setup we have $a=1$ Mm and $b=0.5$ Mm.
The interior region is denser than the exterior, 
and the density, $\rho$, increases over the boundary shell
defined as a function of $\rho_e$, $\rho_i$, $a$, and $b$:
\begin{equation}
\label{densitylayer}
\displaystyle{\rho(\rho_e,\rho_i,a,b)=\rho_e+\left(\frac{\rho_i-\rho_e}{2}\right)\left[1-tanh\left(\frac{e}{a-b}\left[r-\frac{b+a}{2}\right]\right)\right]},
\end{equation}
where $r=\sqrt{x^2+y^2}$ is the radial distance from the centre of the cylinder,
$\rho_e=1.16\times10^{-16}$ g/cm$^{-3}$ is the density in the exterior region,
and $\rho_i$ is the density in the interior where $\rho_i=4\rho_e$.
In this model the initial 
magnetic field, $\vec{B}$, along the $z$-direction
and the thermal pressure, $p$, are uniform,
where the initial magnetic field strength is $B_0=5.6$ G
and the plasma $\beta$ is uniform $\beta=0.02$.
The initial plasma temperature $T$ is set by the equation of state:
\begin{equation}
\label{eos}
\displaystyle{p=\frac{\rho}{0.5 m_p} k_b T}
\end{equation}
where $m_p$ is the proton mass and $k_b$ is the Boltzmann constant
and the initial temperature ranges between $0.34$ (interior) and $1.35$ MK (exterior).

We solve the MHD equations numerically using the MPI-AMRVAC software~\citep{Porth2014},
where thermal conduction, magnetic diffusion, and joule heating are treated as source terms:
\begin{equation}
\label{mass}
\displaystyle{\frac{\partial\rho}{\partial t}+\vec{\nabla}\cdot(\rho\vec{v})=0},
\end{equation}
\begin{equation}
\label{momentum}
\displaystyle{\frac{\partial\rho\vec{v}}{\partial t}+\vec{\nabla}\cdot(\rho\vec{v}\vec{v})
   +\nabla p-\frac{\vec{j}\times\vec{B}}{c}=0},
\end{equation}
\begin{equation}
\label{induction}
\displaystyle{\frac{\partial\vec{B}}{\partial t}-\vec{\nabla}\times(\vec{v}\times\vec{B})=\eta\frac{c^2}{4\pi}\nabla^2\vec{B}},
\end{equation}
\begin{equation}
\label{energy}
\displaystyle{\frac{\partial e}{\partial t}+\vec{\nabla}\cdot[(e+p)\vec{v}]=-\eta j^2-\nabla\cdot\vec{F_c}}, 
\end{equation}
where $t$ is time, $\vec{v}$ velocity,
$\eta$ the magnetic resistivity, $c$ the speed of light, $j=\frac{c}{4\pi}\nabla\times\vec{B}$ the current density, and
$F_c$ the conductive flux \citep{Spitzer1962}.
The total energy density $e$ is given by
\begin{equation}
\label{enercouple}
\displaystyle{e=\frac{p}{\gamma-1}+\frac{1}{2}\rho\vec{v}^2+\frac{\vec{B}^2}{8\pi}},
\end{equation}
where $\gamma=5/3$ denotes the ratio of specific heats.
The numerical experiments presented in this Section 
are ideal and $\eta=0$.

The computational grid has a uniform resolution of $\Delta x=\Delta y=0.15$. Mm
and $\Delta z=0.78$ Mm. 
The simulation domain extends from $z=0$ Mm to $z=200$ Mm in the direction of the initial magnetic field
and horizontally from $x=-2$ Mm to $x=2$ Mm and from $y=-2$ Mm to $y=2$ Mm.
The domain will be enlarged horizontally when a driver composed of 1000 pulses is used, because of the larger spatial displacement of the magnetised cylinder.
The numerical resolution is chosen such that
the long duration of the simulation remains computationally viable.
The boundary conditions are treated with a system of ghost cells,
and we have zero gradient boundary conditions at both $x$ and $y$ boundaries and 
the upper $z$ boundary.
The driver is set as a boundary condition at the lower $z$ boundary.

Here, we use this numerical setup to run four simulations
with four different footpoint drivers (with different periods and amplitudes) that are picked from the pulses used to construct the power spectrum in Sec.\ref{multifdriver}.

\subsection{Electric currents}
\label{idealcurrent}
In these numerical experiments, 
wavetrains propagate into the domain as soon as the driver sets in.
The general dynamics of this kind of system has been discussed in detail previously by \citep{Pascoe2010,PaganoDeMoortel2017} and here we only focus on the 
induction of electric currents.

\begin{figure*}
\centering

\includegraphics[scale=0.23]{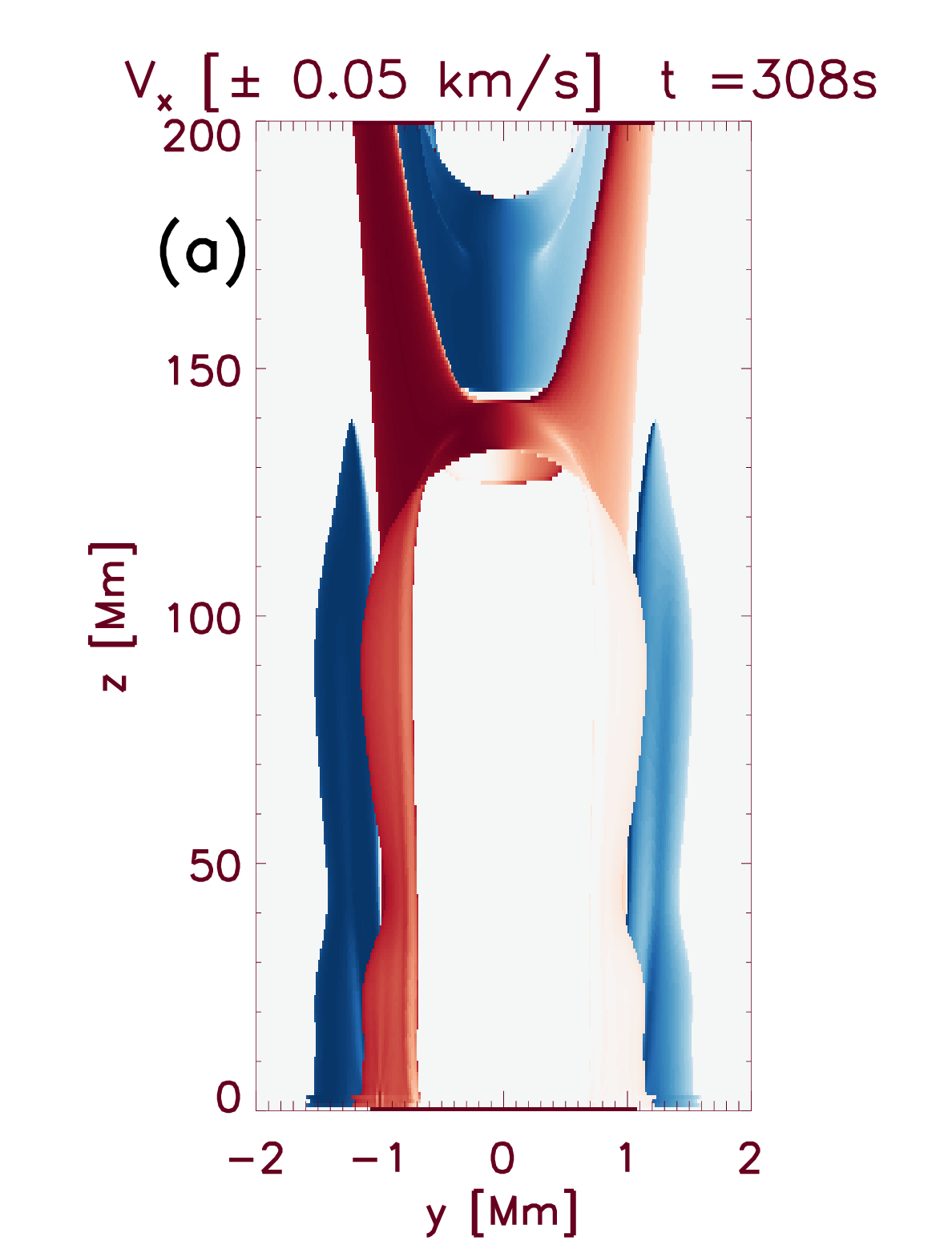}
\includegraphics[scale=0.23]{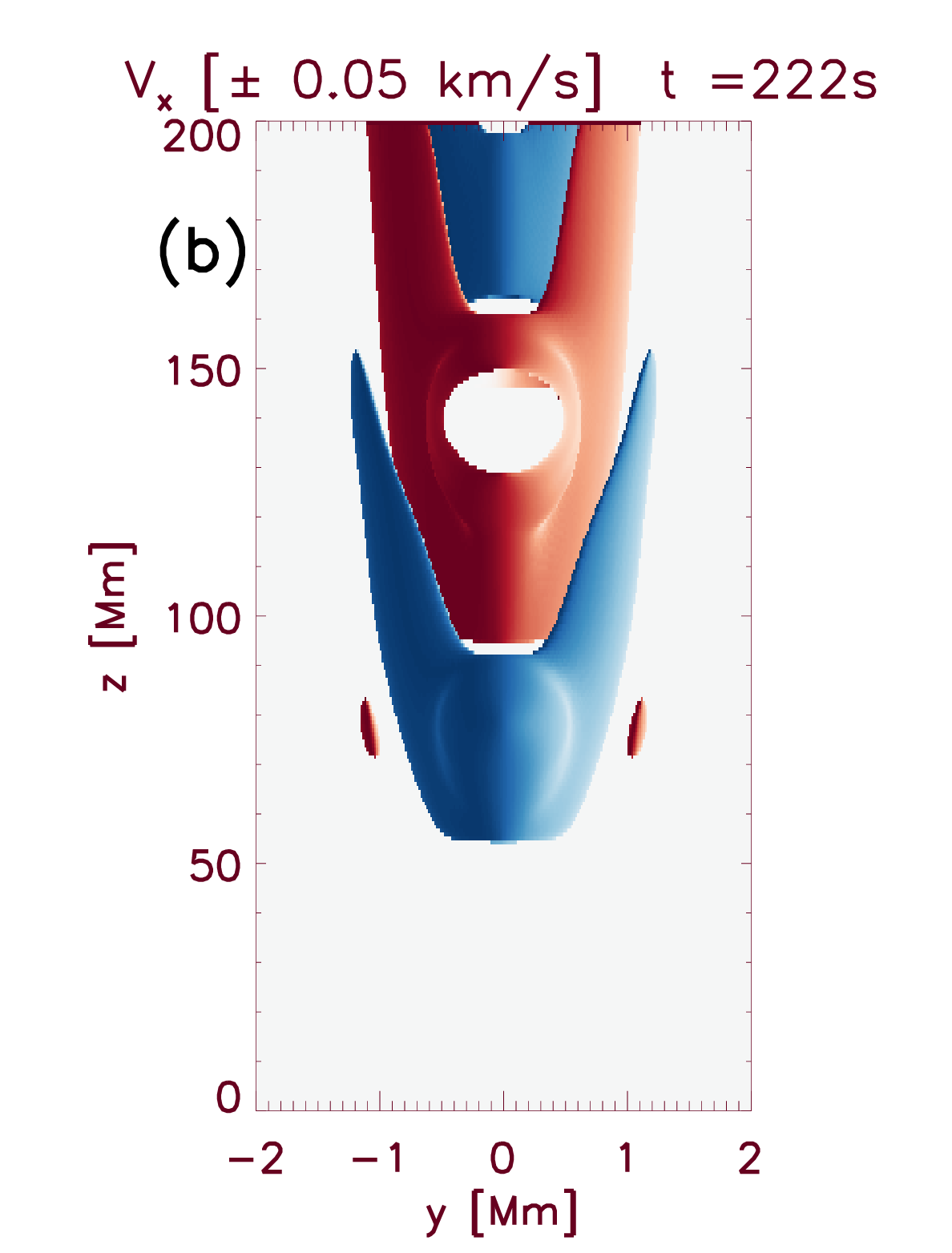}
\includegraphics[scale=0.23]{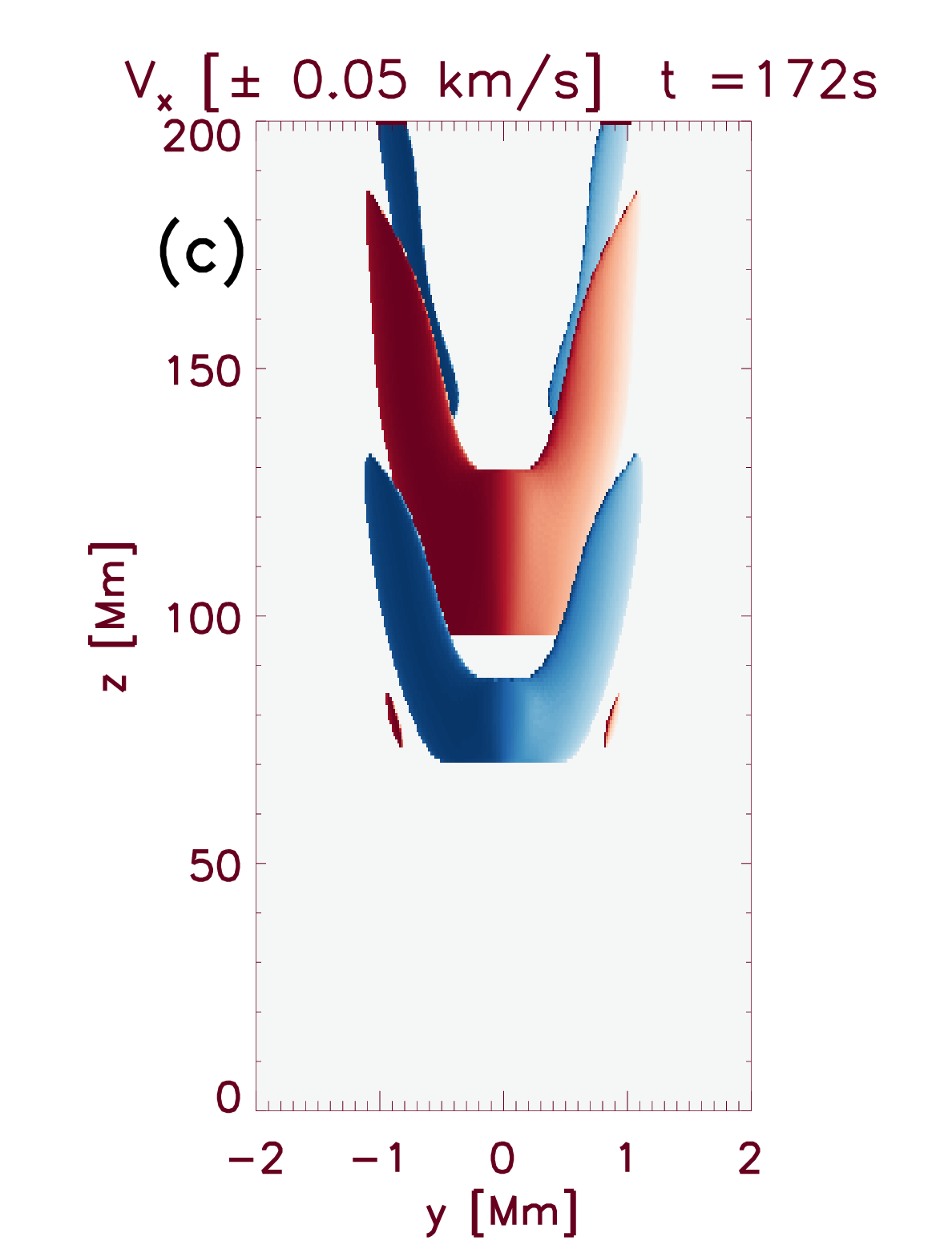}
\includegraphics[scale=0.23]{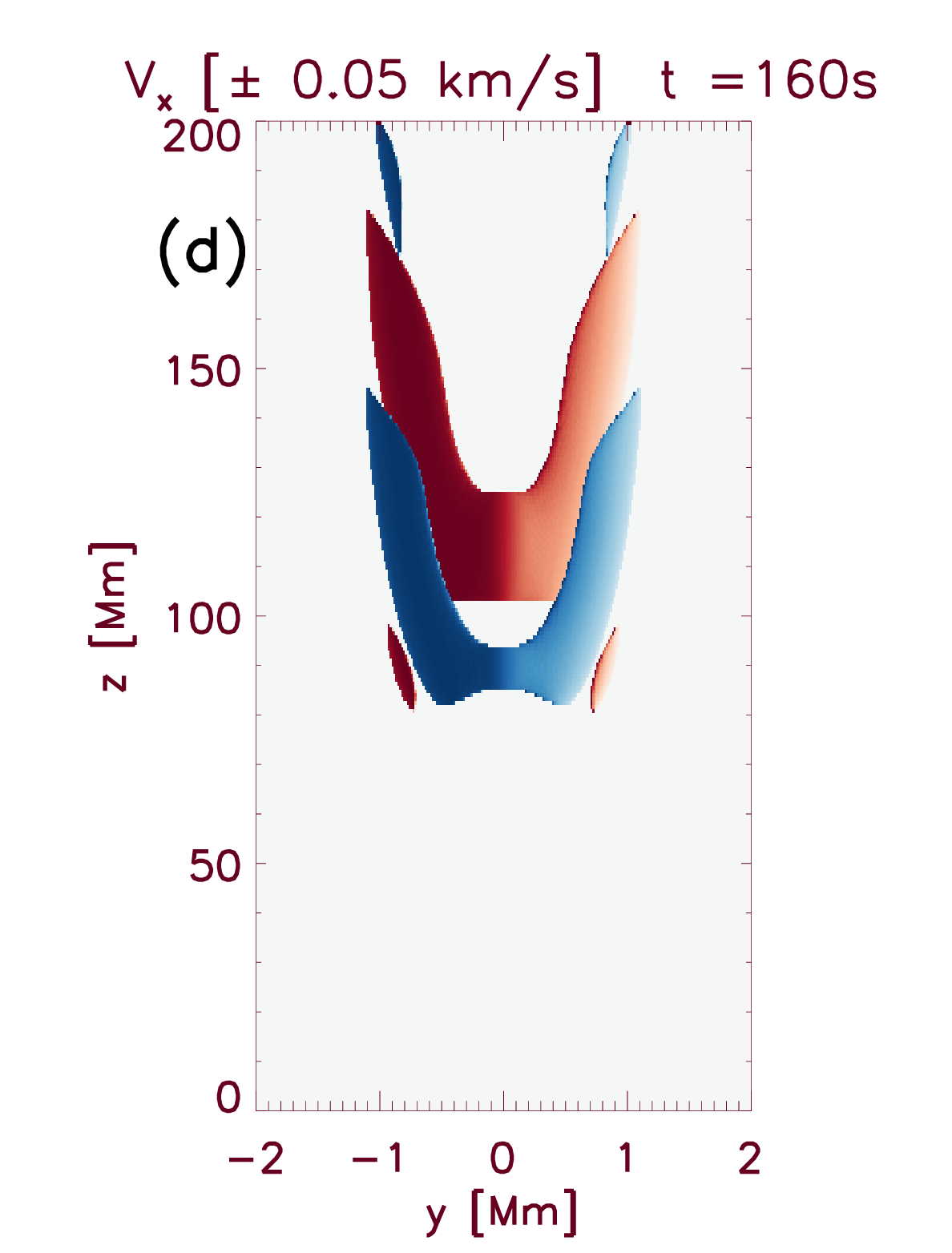}

\includegraphics[scale=0.23]{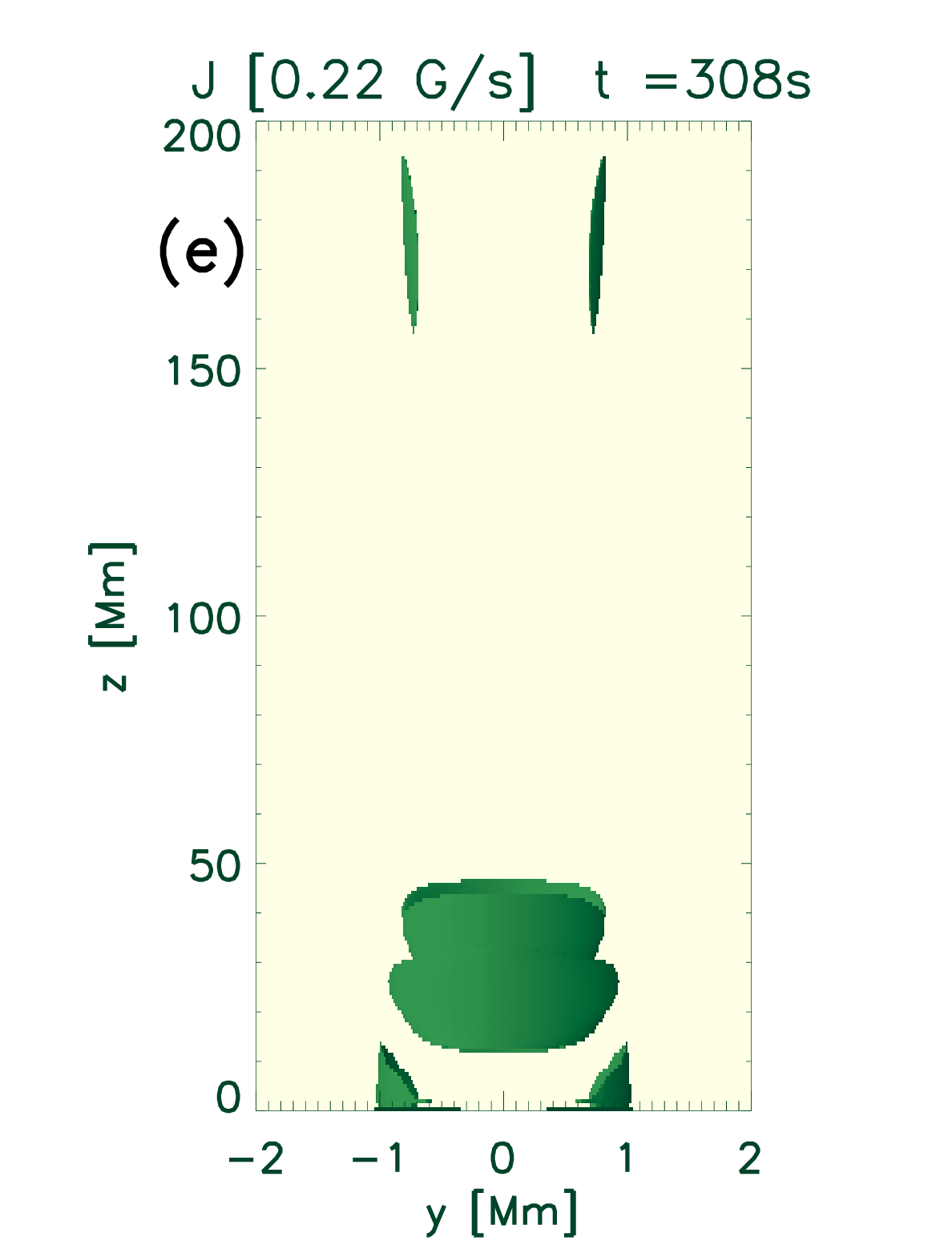}
\includegraphics[scale=0.23]{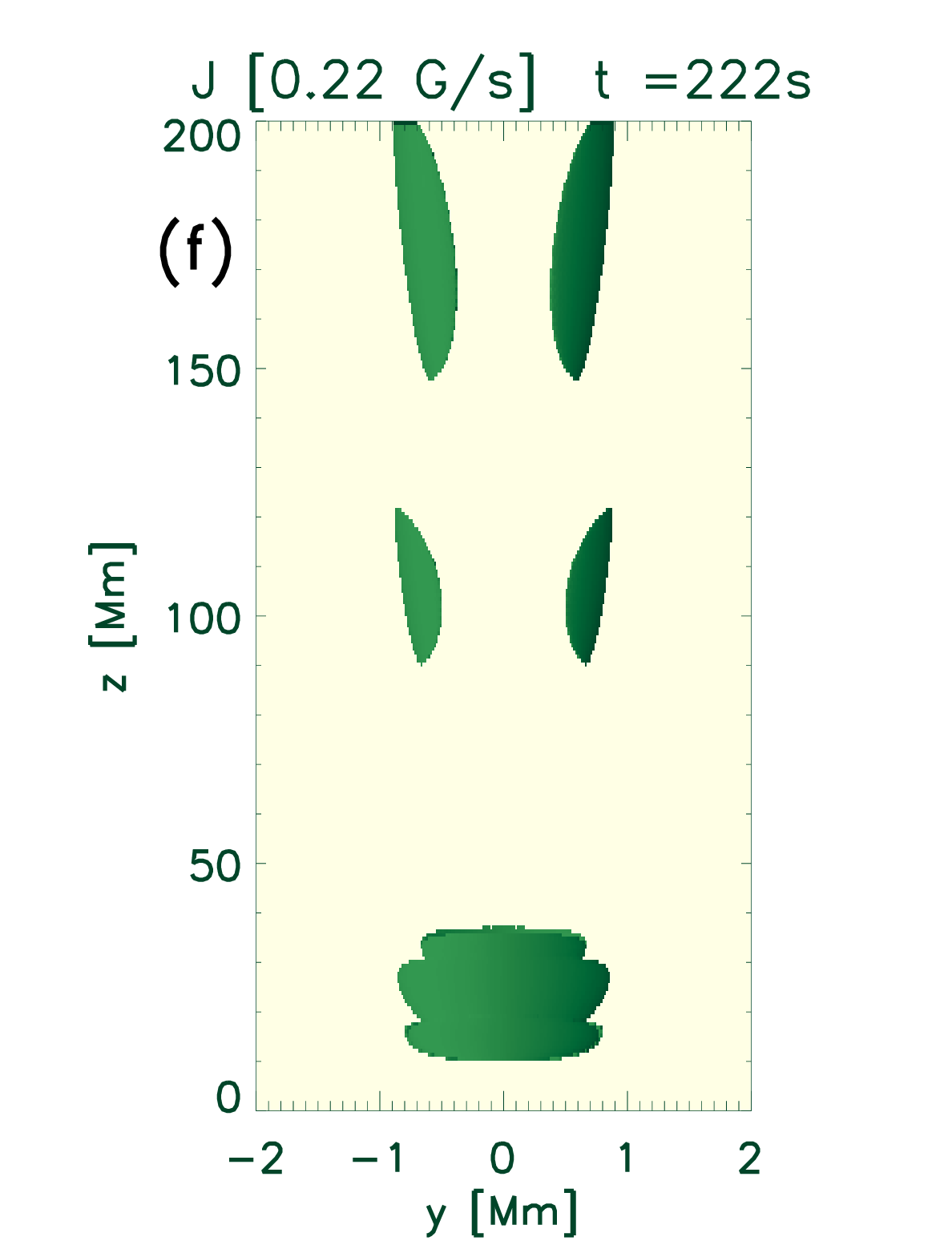}
\includegraphics[scale=0.23]{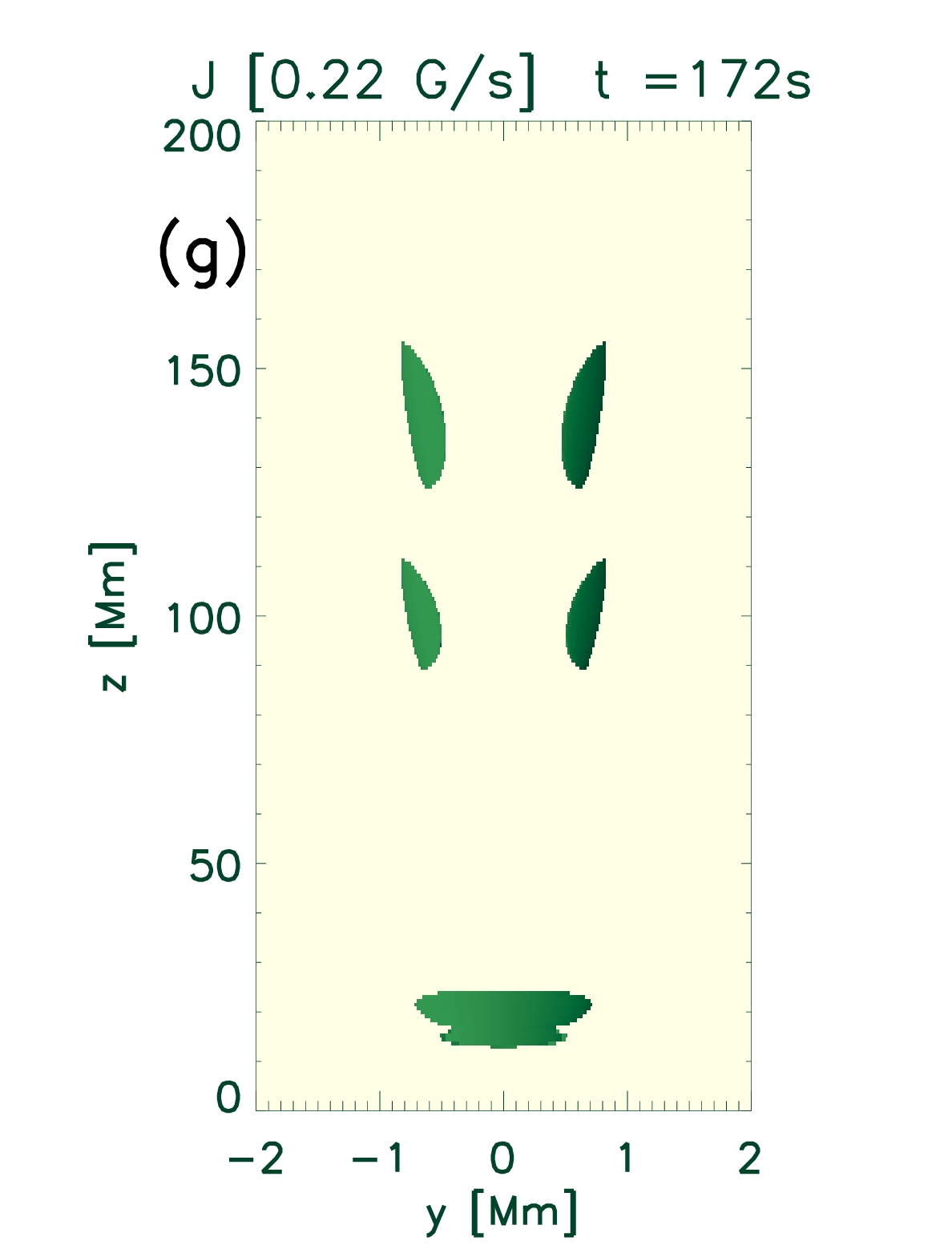}
\includegraphics[scale=0.23]{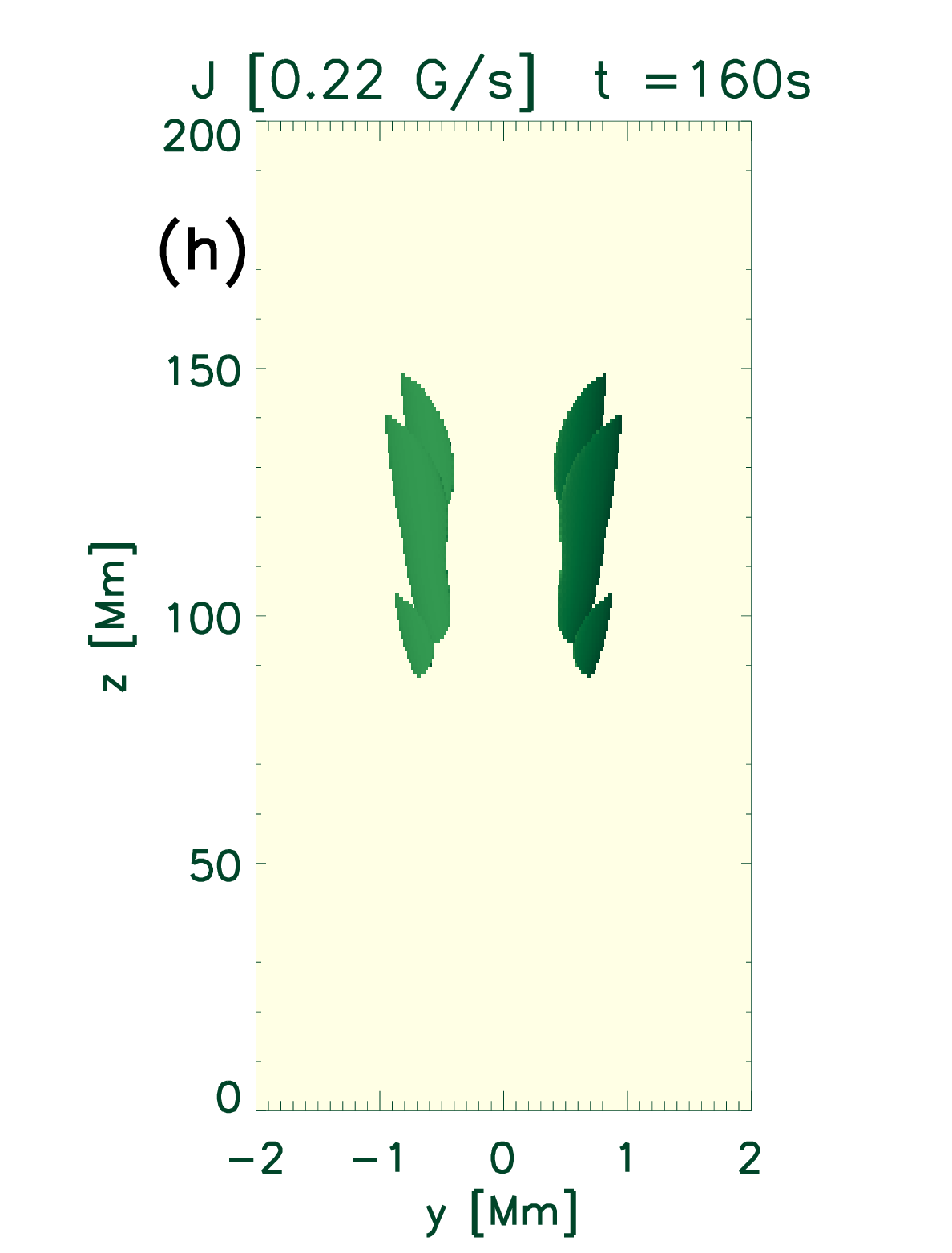}
\caption{a-d 3D contours of $v_x$ ($1$ km/s)
and e-h electric current modulus ($0.22$ G/s)
for the four simulations at different times (from left to right) where
$P=533$ s, $P=178$ s, $P=107$ s, and $P=76$ s.
For $v_x$ we represent positive velocities in blue and negative in red.}
\label{currentssim}
\end{figure*}
We illustrate the evolution of velocity and currents in 
Fig. \ref{currentssim}, where we show
3D contours of velocity (panels a-d)
and electric currents (panels e-h) for the four simulations 
at comparable stages of their evolution.

The velocity contours (Fig.\ref{currentssim}a-d) show that the phase-mixing is more advanced 
for the simulations with the shorter period drivers, as the elongated 
velocity structures show that different Alfv\'en waves have travelled significantly out of phase.
This is reflected in the different electric current distributions.
In Fig.\ref{currentssim}e-h lower row, we show the 3D contour of $|\vec{j}|=0.22$ G/s. As expected, we find that similar levels of current are induced sooner (i.e.~at lower values of $z$) for shorter period drivers. This is because the longer-period Alfv\'en waves need to travel longer to reach the same phase-mixing stage. We also see that the simulations with longer period driver show the induction of electric current propagating upwards at the local acoustic speed near the (driven) footpoint.
These currents are generated 
by slow-mode waves propagating upwards following the compression of the guide field due to the displacement of the loop,
where this effect is larger for the long-period waves due to
their larger amplitudes.
Finally, at the lower boundary, the shear motion between the magnetised cylinder and the surrounding background where the driver is not imposed 
generates additional currents that are weak and localised.

\begin{figure}
\centering
\includegraphics[scale=0.28]{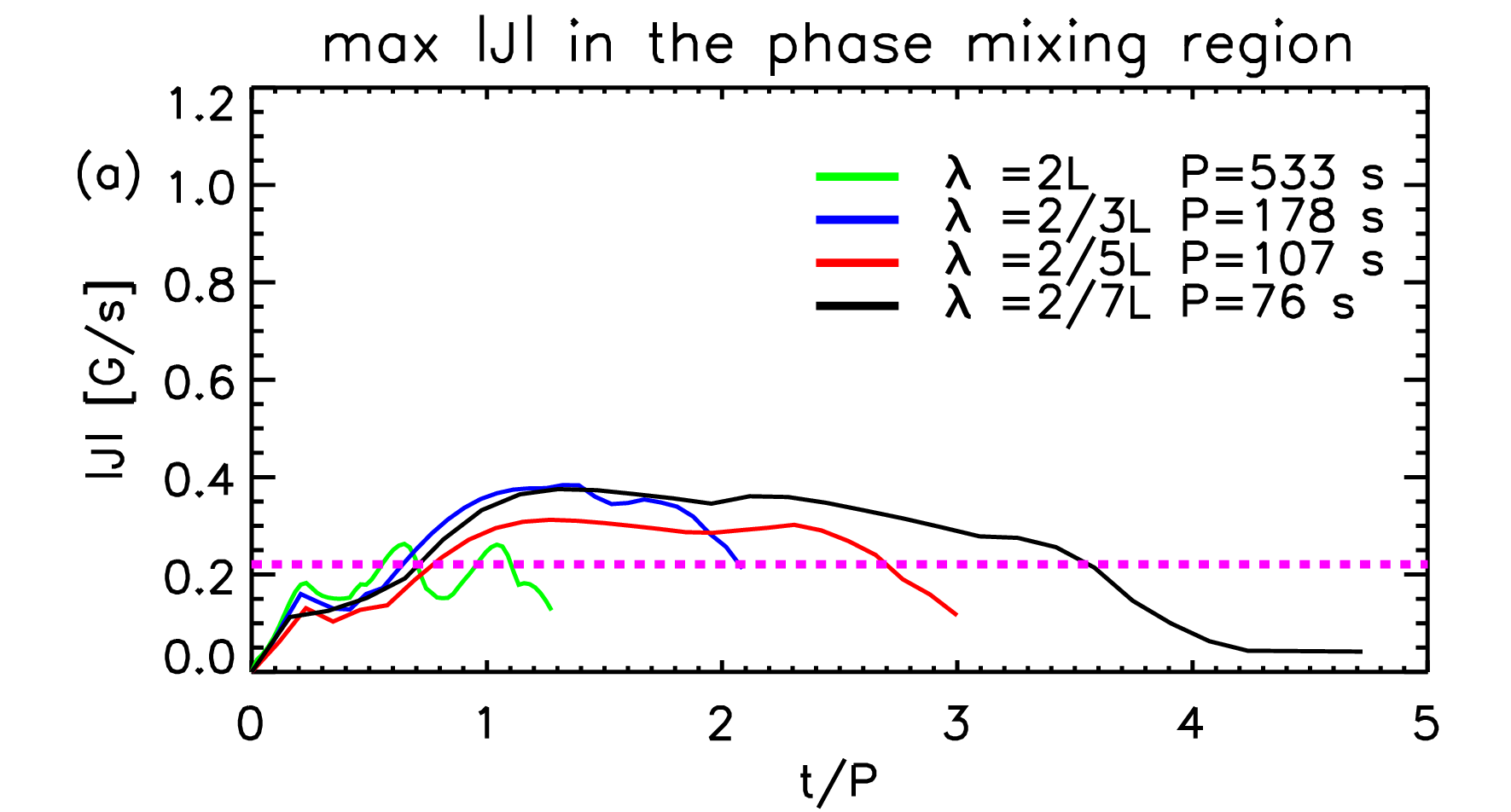}
\includegraphics[scale=0.28]{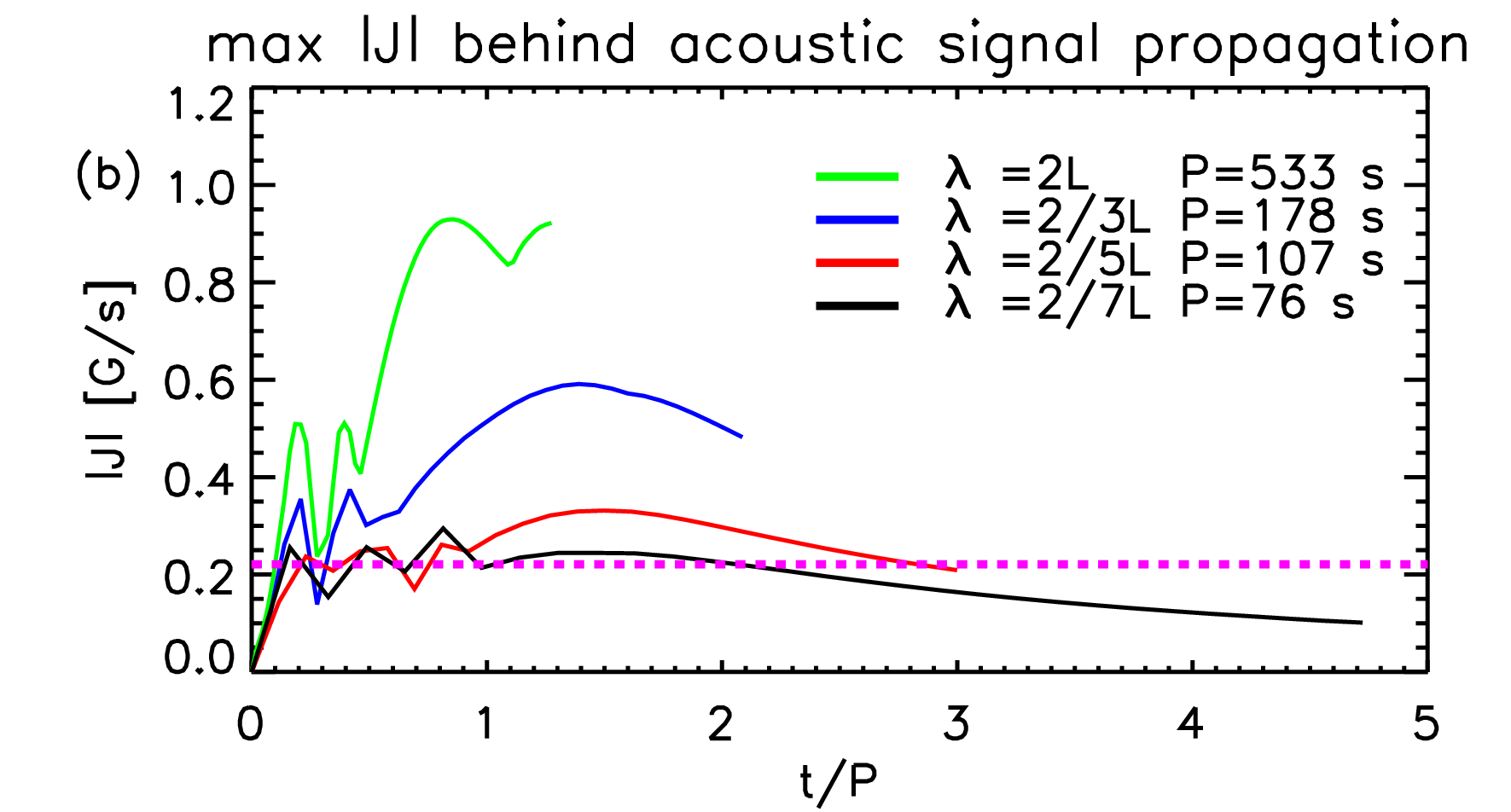}
\caption{For the fours simulations where
$P=533$ s, $P=178$ s, $P=107$ s, $P=76$ s
we show 
(a) the maximum electric current modulus as a function of time (normalised to the period of the Alfv\'en wave)
in the region beyond the z coordinated reached by a sound wave at each time
and (b) for the region below the same location.}
\label{currentssimtime}
\end{figure}
Fig.\ref{currentssimtime}a shows the maximum current modulus in the region above the z-coordinate reached by a slow-mode signal,
in order to represent only currents induced by phase-mixing.
We find that for all simulations the current modulus initially increases and then drops when the waves leave the domain.
All simulations that show significant phase-mixing 
eventually develop current in this regime and the phase-mixing 
sustains these currents throughout the simulations. 
Fig.\ref{currentssimtime}b shows instead
the maximum current modulus in the region below the z coordinate reached by a slow-mode signal,
that excludes phase-mixing induced currents.
This time, the strongest currents are found when the driver has higher velocity amplitude.
These currents are induced by the slow-modes propagating upwards that are a physical feature connected to
the displacement of a dense magnetised cylinder while the background corona remains at rest.
This scenario is not just an artefact of our simulations, as such
electric currents are expected to be induced in the solar corona 
when loop foot points are set in motion by photospheric flows.

This set of numerical experiments shows that 
we can expect two types of electric current when 
the footpoint of a magnetised cylinder is displaced.
On one hand, we have currents propagating at the acoustic speed that are
induced near the footpoint by the compression of the guide magnetic field and hence are proportional to the velocity amplitude.
On the other hand, we have currents generated by the phase-mixing of 
propagating Alfv\'en waves that depend on the driver frequency.
Moreover, this study allows us to identify a threshold current ($|\vec{j}|=0.22$ G/s) for this particular numerical experiment that selects fully developed phase-mixing currents.
This threshold value is an ad-hoc choice that satisfies our modelling requirements but does not bear a general physical meaning.
The threshold has been selected to allow dissipating all the phase-mixing induced currents and it is thus lower than all such currents (Fig.\ref{currentssimtime}a).
It should be noticed however that Fig.\ref{currentssimtime} shows the maximum currents in the domain and at any time some regions indeed show currents larger than the threshold. At the same time, the current threshold is larger than most of the currents in the domain and no physical dissipation occurs in these regions.

\subsection{Plasma thermodynamics}
\label{idealsimulation}
In addition to the electric currents induced during 
the wave propagation, 
it is key to understand the plasma thermodynamic evolution
in this ideal MHD numerical experiment
in order to assess the plasma heating due to the dissipation of 
currents when the magnetic resistivity is effective.

\begin{figure}
\centering
\includegraphics[scale=0.36,clip,viewport=155 65 500 640]{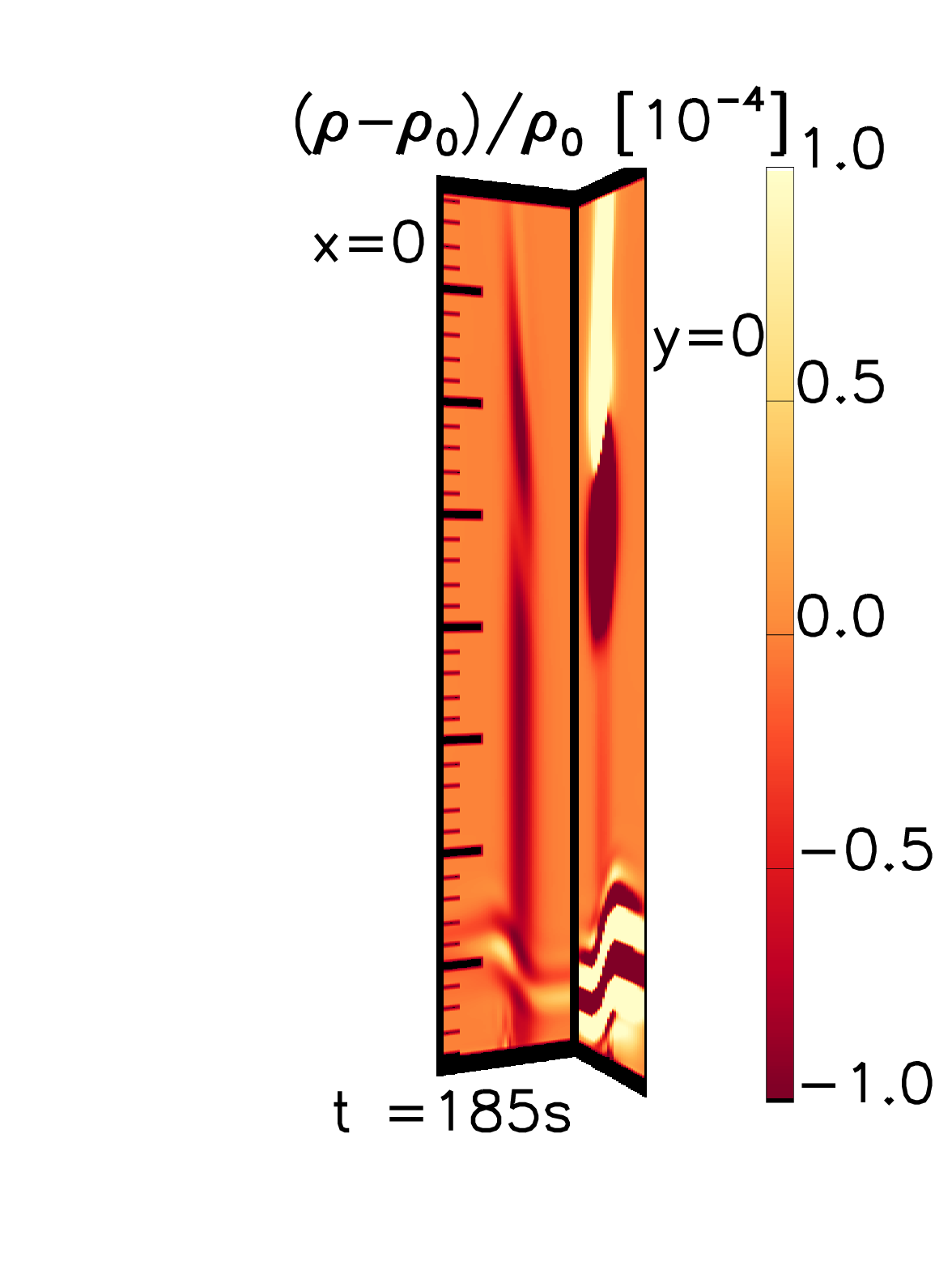}
\includegraphics[scale=0.36,clip,viewport=155 65 500 640]{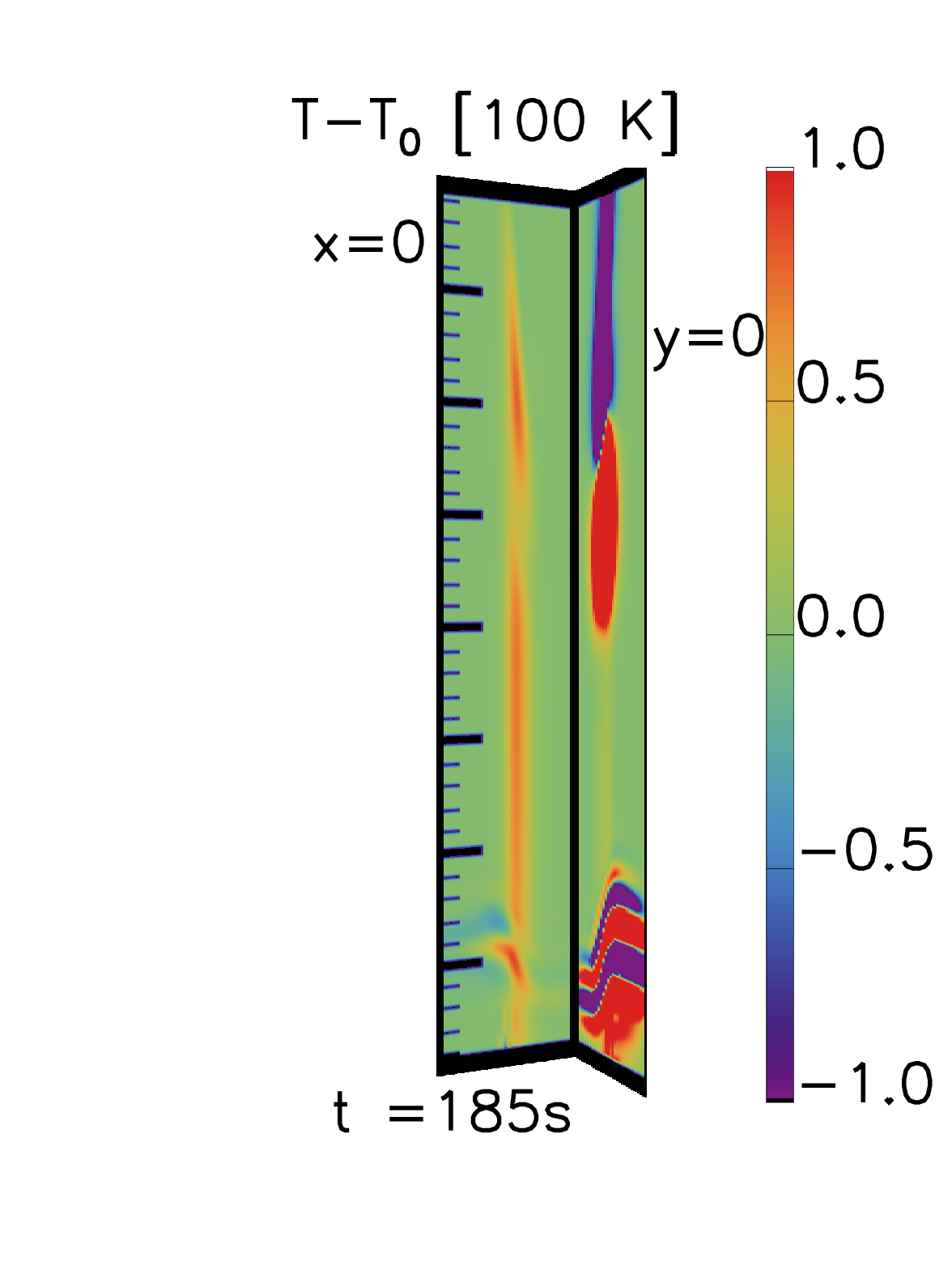}

\caption{3D cuts of the MHD simulation domain showing maps of
density relative change (a) and temperature change (b) at $t=185$ s for the simulation with $P=76$ s.}
\label{rhotemp35}
\end{figure}

Here, we analyse only the simulation where $P=76$ s
as a representative case.
Fig.\ref{rhotemp35}a shows the density and
Fig.\ref{rhotemp35}b the temperature change
at $t=185$ s with respect to $t=0$.
At this time the wave induced by the driver has reached beyond $z=100$ Mm
and the density variations clearly show a region where the plasma has became denser and a region where it became less dense above this surface
on the $y=0$ plane, caused by the weak compression and rarefaction
associated with the kink modes.
On the $x=0$ plane, where the phase-mixing manifests itself,
we find an elongated structure of lower density and higher temperature with respect to the initial conditions.
The density decrease is very small, as the driver
induces an almost incompressible perturbation,
having an amplitude that is $\sim10^{-4}$ $V_A$
and the temperature increase is also $\sim10^{-4}$ times the initial temperature.
This modest change in temperature is due to the readjustment
of the thermal pressure after the pulses travels along the waveguide
and it is not connected to (numerical) dissipation.

\begin{figure}
\centering
\includegraphics[scale=0.20,clip,viewport=40 20 1250 1050]{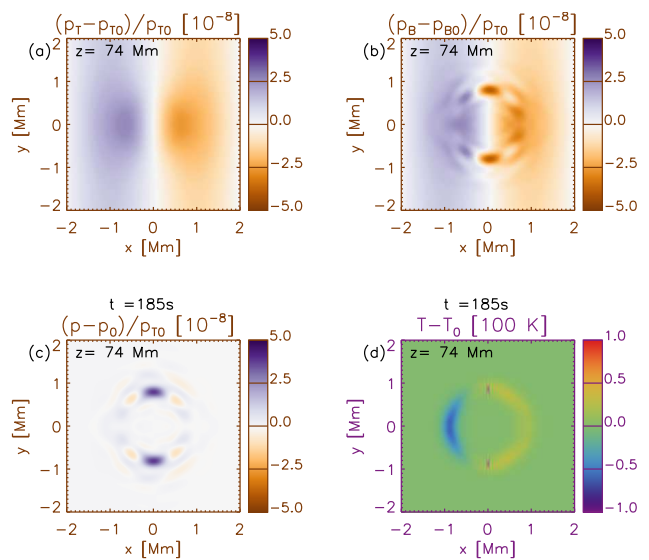} 
\caption{Maps of relative change of (a) total pressure, (b) magnetic pressure, (c) thermal pressure,
and (d) absolute change of temperature at the surface $z=74$ Mm at $t=185$ s
for the simulation with $P=76$ s.}
\label{presmodes35}
\end{figure}

Fig.\ref{presmodes35} shows cross sections at $z=74$ Mm
at $t=185$ s (just after the pulse has crossed this location)
of  the relative $p_{tot}$ variation,
and the variations of $p$ and $p_{mag}$
with respect to the initial total pressure,
and of the temperature variation.
We find that for the total pressure, these variations
are smaller than $10^{-8}$ times the initial value (Fig.\ref{presmodes35}a) as they are the result of compensating adjustments of the magnetic pressure (Fig.\ref{presmodes35}b) and thermal pressure (Fig.\ref{presmodes35}c).
In particular, we find that this reshuffle leads to an increase in thermal pressure on the
boundary shell at the $x=0$ locations.
This is also where the temperature increase is most evident (Fig.\ref{presmodes35}d),
as the temperature is altered only in the boundary shell by the passage of the pulse.
These small changes in density and temperature are the dominating dynamics in this simulation
which we will use as our reference to measure the role of resistivity.

\subsection{Role of magnetic resistivity}
Having investigated the thermodynamic effects of a pulse propagating along the magnetic cylinder
in an ideal MHD regime,  we now study how magnetic resistivity 
can lead to the dissipation of the associated wave energy into heating.

In such a numerical experiment, we need to meet two competing requirements.
On one hand, we need the dissipation of waves to be sufficiently efficient to convert a noticeable amount of energy into heating. On the other hand, we need waves to propagate sufficiently to phase-mix.
At the same time, our numerical simulations are aimed at studying the heat deposition from Alfv\'en waves,
but it is beyond the scope of this work to study the details of this mechanism.
Therefore, an anomalous magnetic resistivity allows for the quick dissipation of 
electric currents generated by the phase-mixing and to study the effects of the subsequent thermal energy deposition.
We adopt an anomalous resistivity of the form
\begin{equation}
\eta=\eta_0\left(\frac{|\vec{j}|^2}{j_0^2} +1\right)
\hspace{1 cm}
\left(|\vec{j}| > j_0\right)
\label{resistivity}
\end{equation}
where we use $j_0=0.22$ G/s as the threshold current and
$\eta_0$ is a multiple of $\eta_S$,
the value of the magnetic resistivity according to \citet{Spitzer1962} at $T=2$ MK.
The value of $j_0$ is conveniently chosen as it is lower than the electric currents 
generated by the phase-mixing of the short-period waves, but 
higher than the electric currents generated by the slow modes associated with
most of the long-period oscillations.
Additionally, we run a series of MHD simulations where
we solve the non-ideal MHD equations
(Eq.\ref{mass}-\ref{energy}) and vary the value of $\eta_0$
to find that $\eta_0=10^{10}\eta_S$ is a value that allows for the effective dissipation of phase-mixing currents.
To conclude, this set of values for $\eta_0$ and $j_0$ ensures that
the wave energy is mostly dissipated once the phase-mixing sets in, whilst ensuring that the dissipation is not instantaneous so that phase-mixing is allowed to develop.

\section{Solar Simulation}
\label{solarsimulation}

Now we have run a set of numerical experiments with single pulse drivers,
we continue with a numerical experiment where the driver is the combined one we have constructed in Sec.\ref{multifdriver}, namely the interference of 1000 pulses with different periods, direction of oscillation and starting time.
In this numerical experiment we use the same initial conditions as in Sec.\ref{simulationsetup} and a value of resistivity of $\eta=10^{10}\eta_S$.
This combined driver leads to a larger spatial displacement of the magnetic cylinder than previously analysed,
thus we take a larger domain in $x$ and $y$
(from $x=-3$ Mm to $x=3$ Mm and from $y=-3$ Mm to $y=3$ Mm).

\begin{figure}
\centering

\includegraphics[scale=0.23]{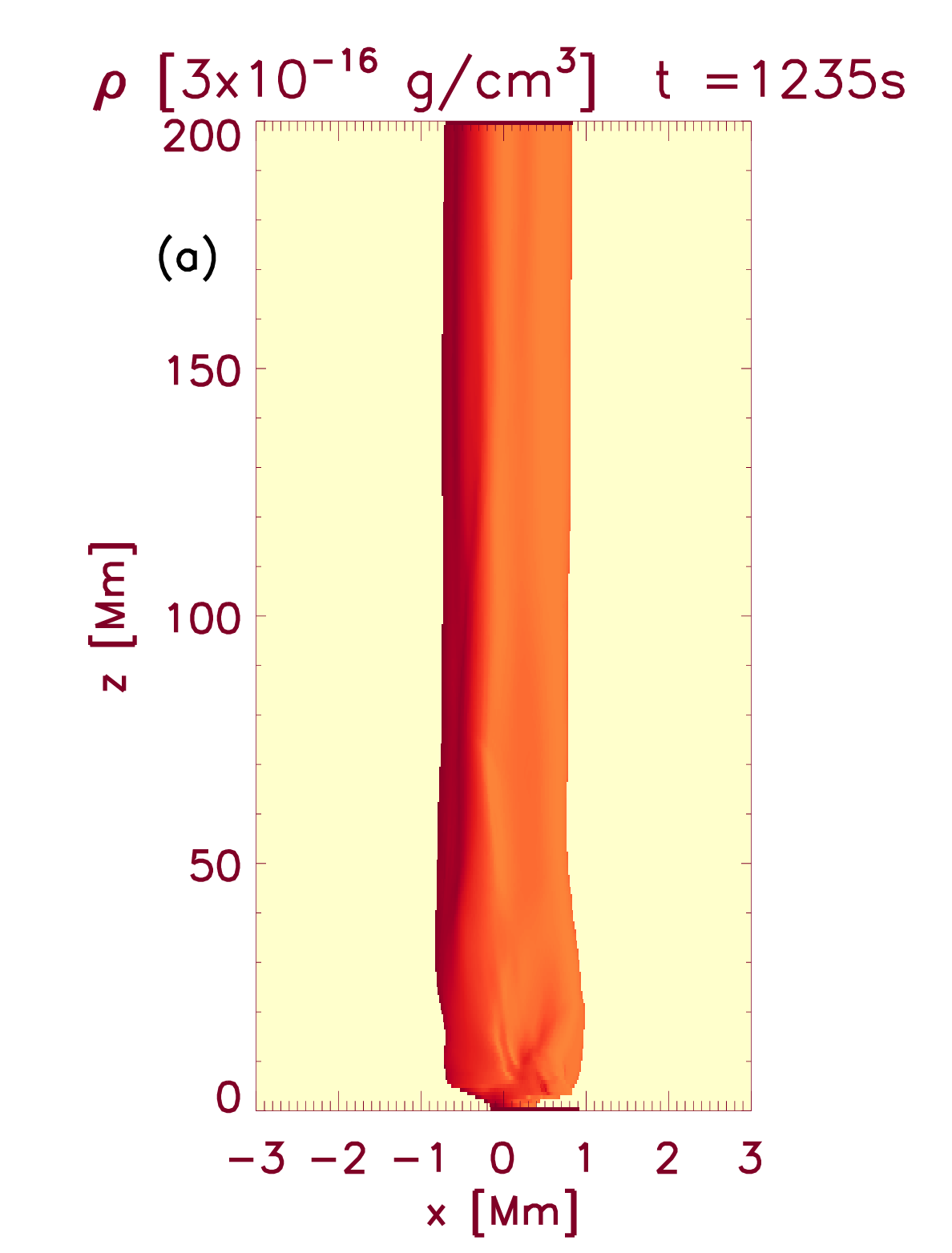} 
\includegraphics[scale=0.23]{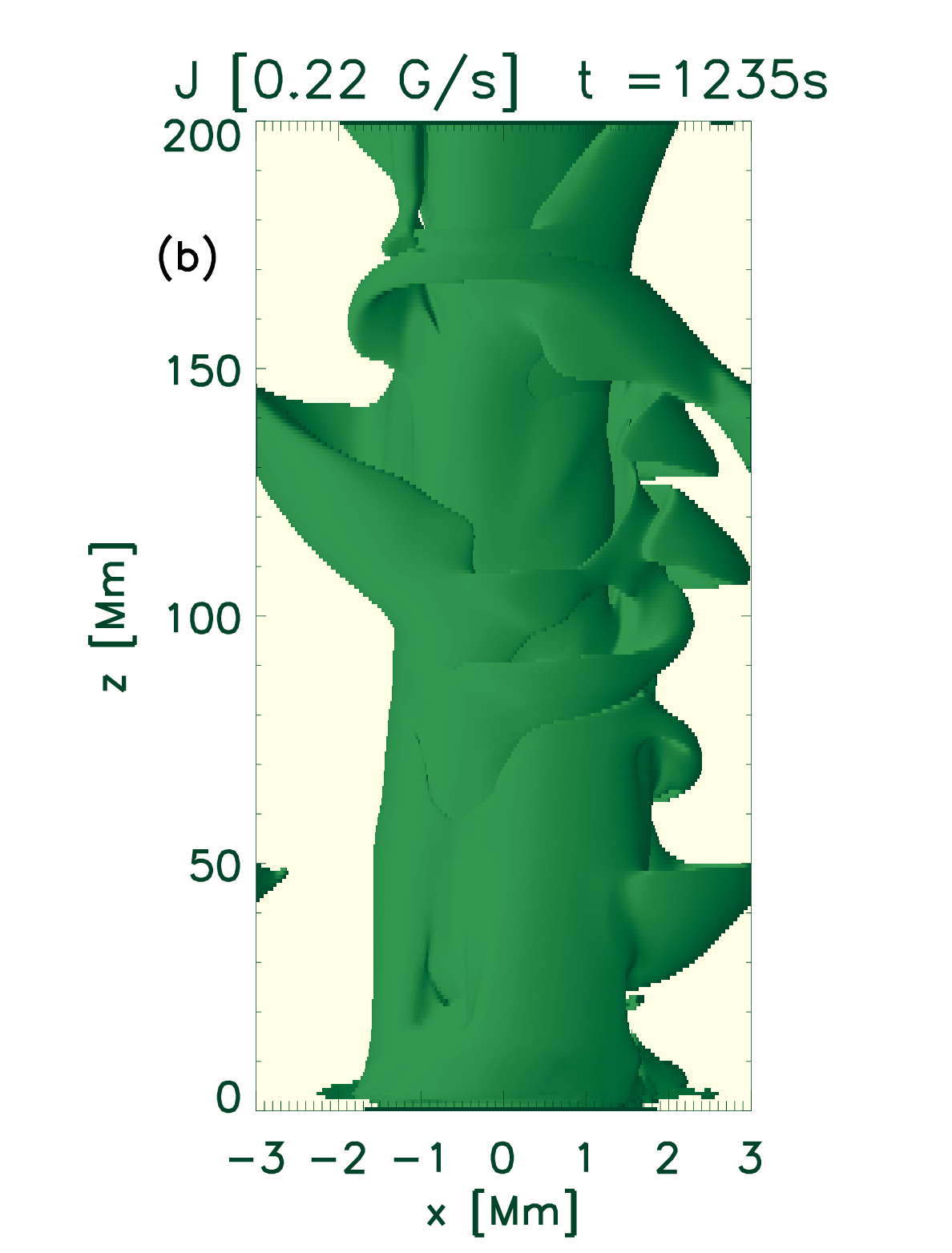} 

\includegraphics[scale=0.23]{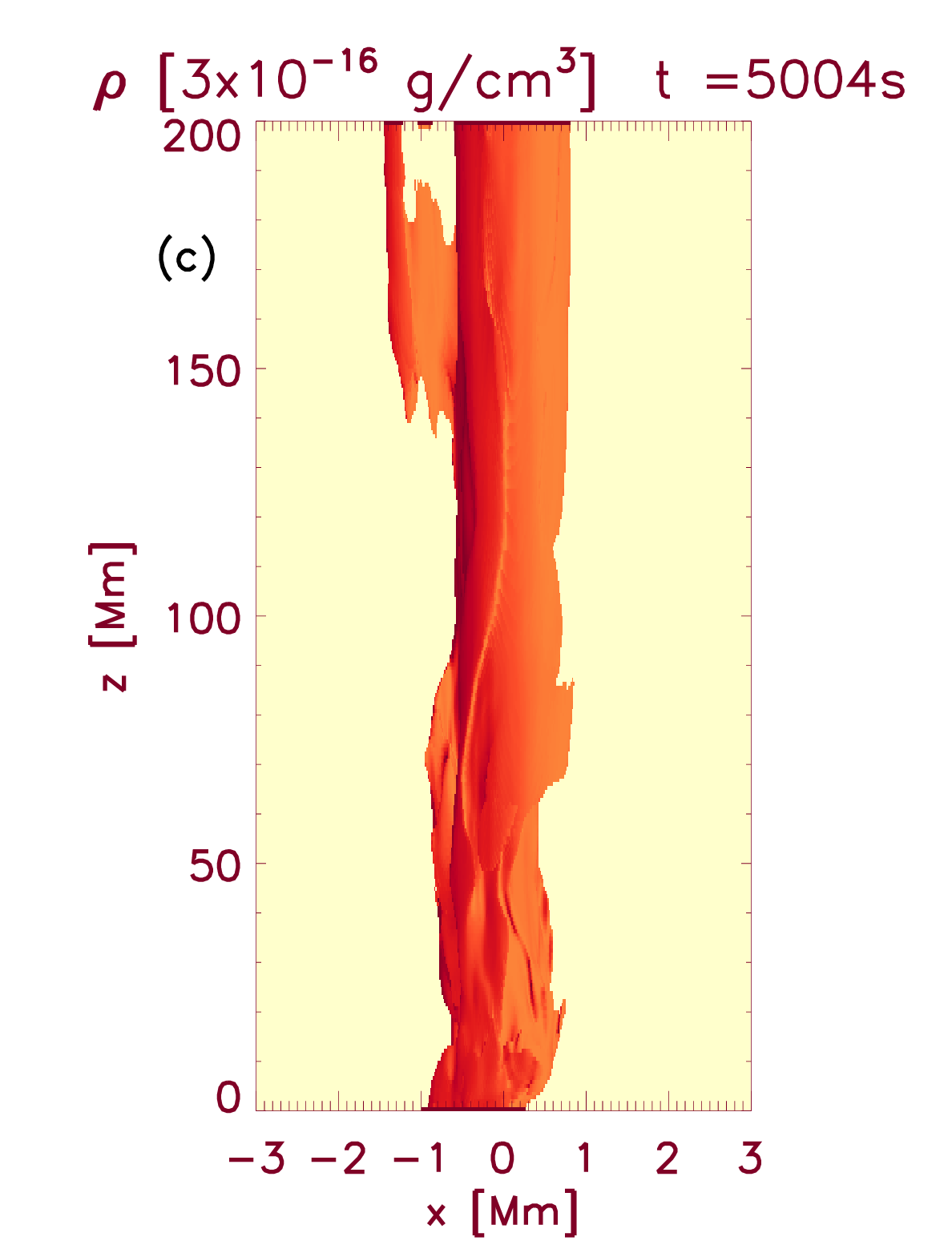} 
\includegraphics[scale=0.23]{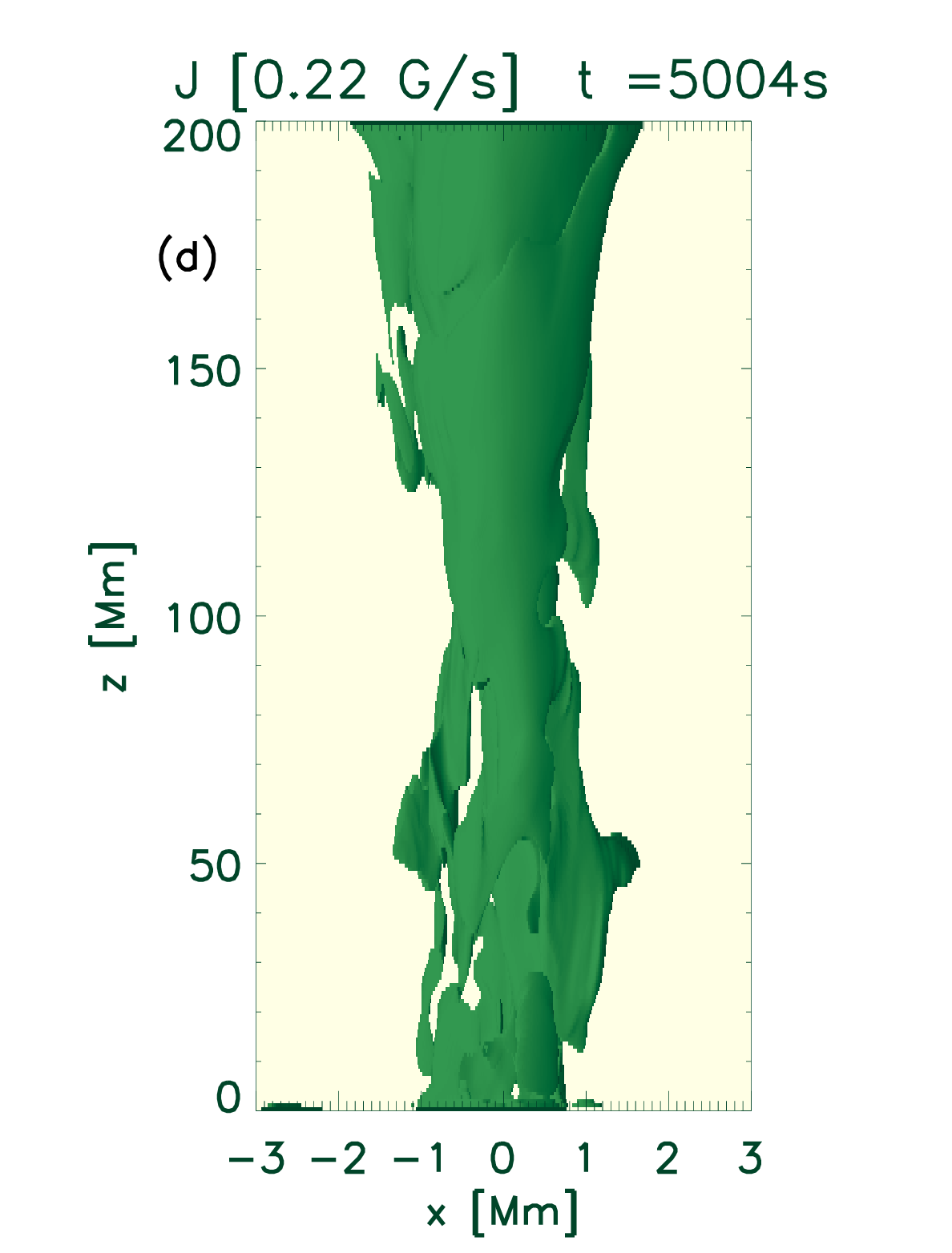} 

\caption{3D contours of density
($3\times10^{-16}$ g/cm$^{-3}$, initial boundary shell density)
and electric current modulus
($j_0=0.22$ G/s)
at $t=1235$ s and at $t=5004$ s.
The temporal evolution is available in a movie online.}
\label{solspedensj}
\end{figure}

This simulation runs for $5000$ s of physical time, which is a time long enough for all the pulses of any period ($1$ $min$ $<$ $P$ $<$ $26$ $min$) to start and finish.
Fig.\ref{solspedensj} shows the 3D contour of density 
and modulus of electric currents at 
$t=1236$ s and $t=5000$ s.
The persistence of the driver and the interference between the various pulses significantly affect the loop structure and thermodynamics.
The amplitude of the displacement is large enough to lead to a visible drift of the loop footpoint, thus the propagation of waves along the magnetised cylinder starts from different location in the $x-y$ lower boundary plane and the density structure is significantly affected in time.
While the structure of a cylinder is still maintained after $1000$ s of evolution, where significant alterations are visible at the footpoint only, the density structure at $5000$ s has undergone major evolution where the cylinder is deformed and fragmented in some parts.
The analysis carried out in Sec.\ref{mhdevolution} suggests that the development of currents above the resistivity threshold is the precursor of plasma heating. We find that the currents above the threshold $j_0$ are present across the entire loop structure and in time, the current distribution becomes more and more fragmented into small-scale structures. 

\begin{figure}
\centering
\includegraphics[scale=0.23]{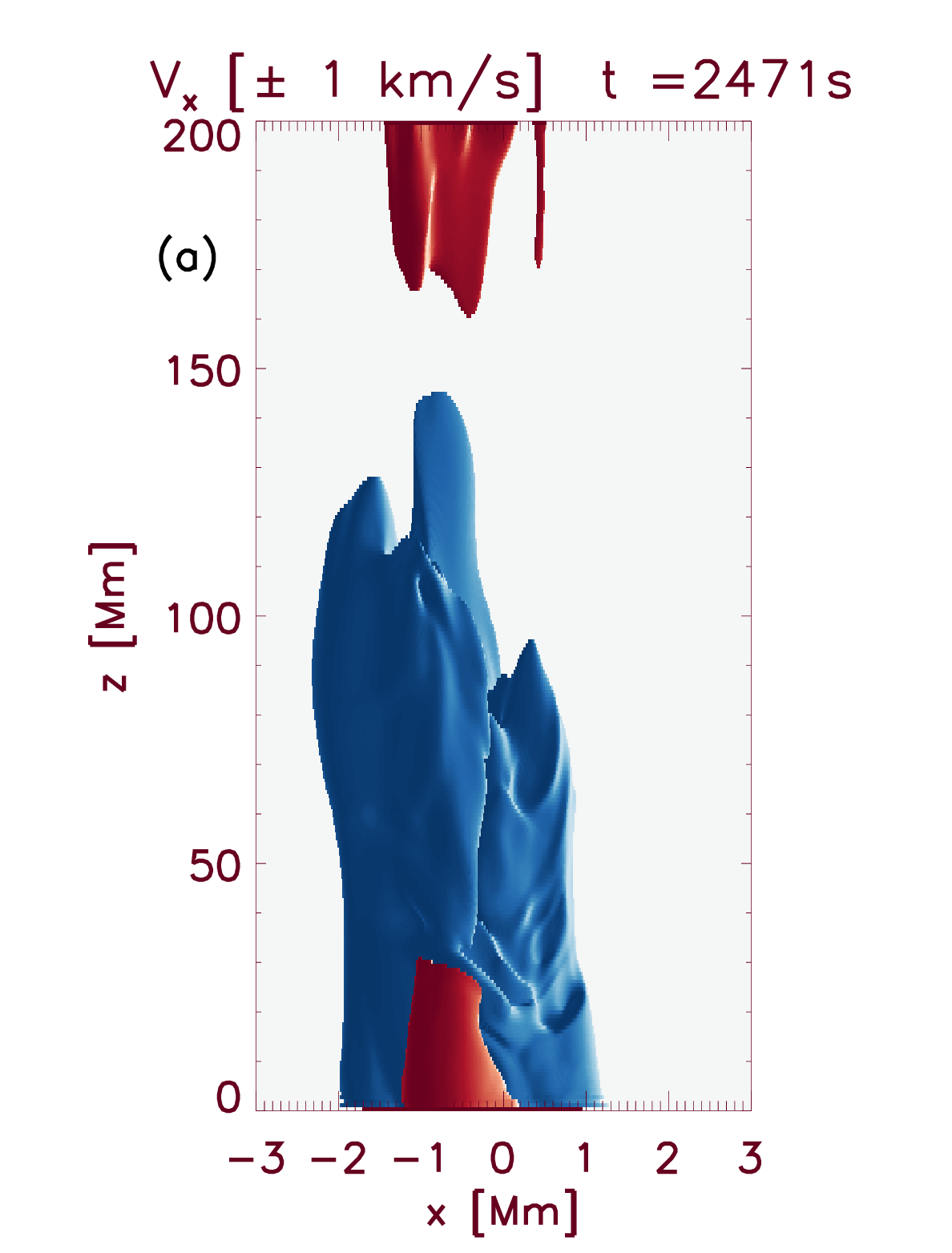}
\includegraphics[scale=0.23]{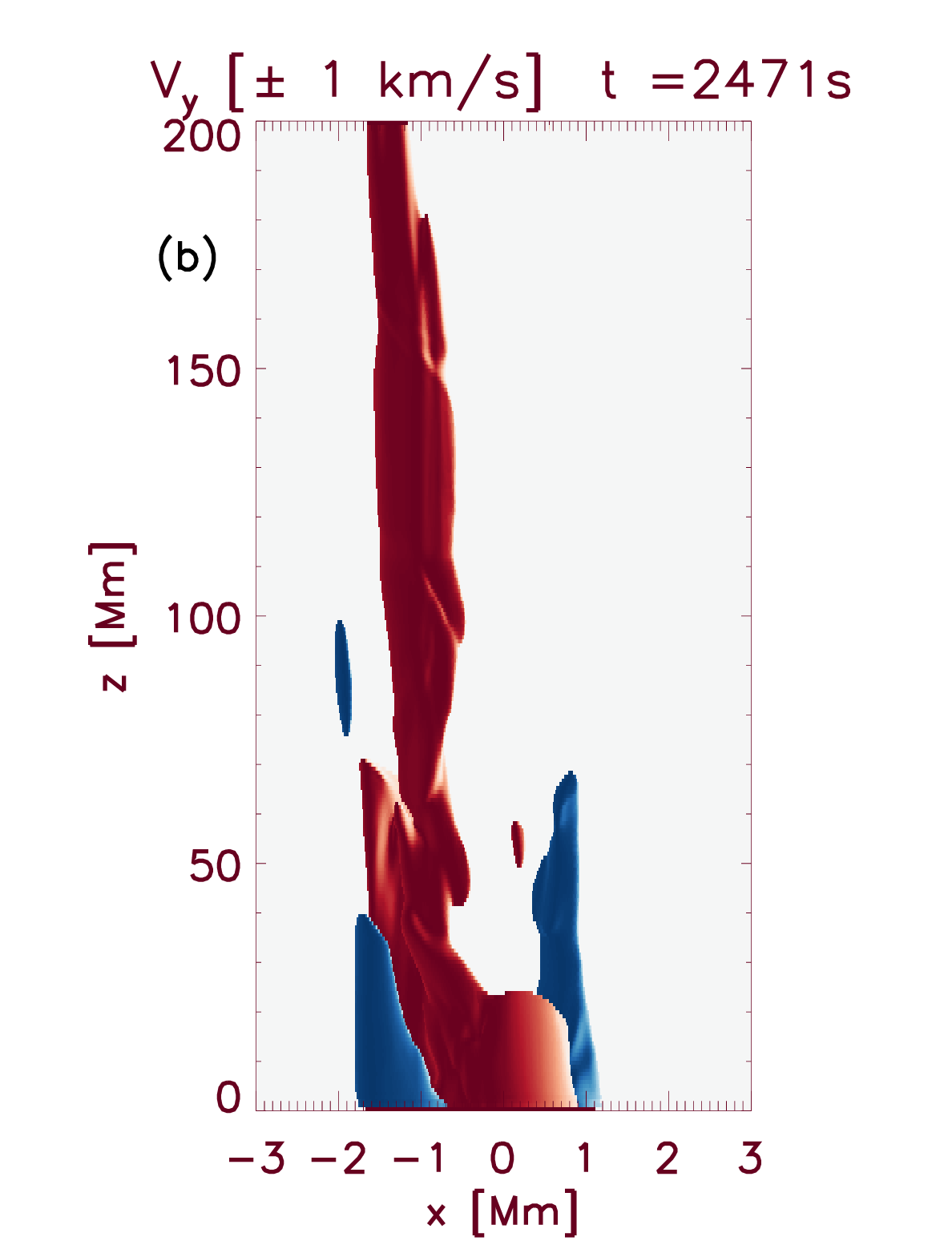}
\includegraphics[scale=0.23]{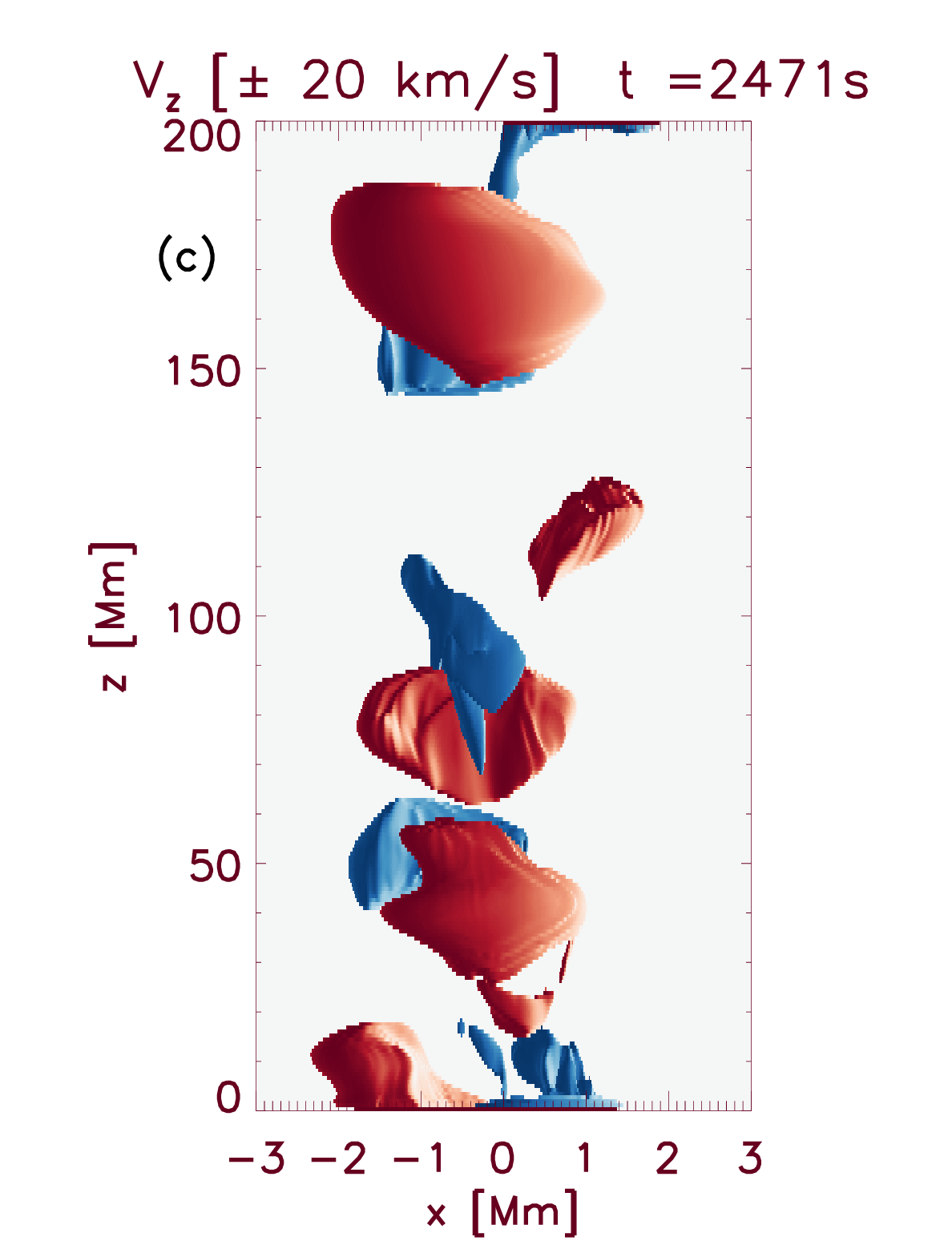}

\caption{3D contours of the velocity components at $t=2471$ s.
We show contours where the absolute value of the velocity components are $1$ km/s for $V_x$ (a) and $V_y$ (b)
and $20$ km/s for $V_z$ (c) and we identify negative velocities with red contours and positive ones with blue contours.
The temporal evolution is available in a movie online.}
\label{solspevxyz}
\end{figure}
Fig.\ref{solspevxyz} shows the 3D contours for the 3 components of the velocity at $t=2471$ s, at which time significant velocities are still present in the domain.
We find that
the distribution of $V_x$ and $V_y$ is what is expected by the wave propagation in the domain triggered by the footpoint motion.
Both transverse velocity components are of the order of $1$ km/s and their distribution presents the usually observed pattern of phase-mixing 
where elongated structures are visible at the boundary shell. 
In contrast, the z-component of the velocity (Fig.\ref{solspevxyz}(c)) is significantly larger, of the order of $10$ km/s. These field-aligned perturbations are generated at the footpoint by the displacement of the loop that compresses the guide field, leading to slow-mode waves propagating upward.
These sporadic, isolated, and slowly propagating flows do not significantly interact or alter the propagation of the fast propagating transverse MHD waves, which are present in a large portion of the domain.

\begin{figure}
\centering
\includegraphics[scale=0.32]{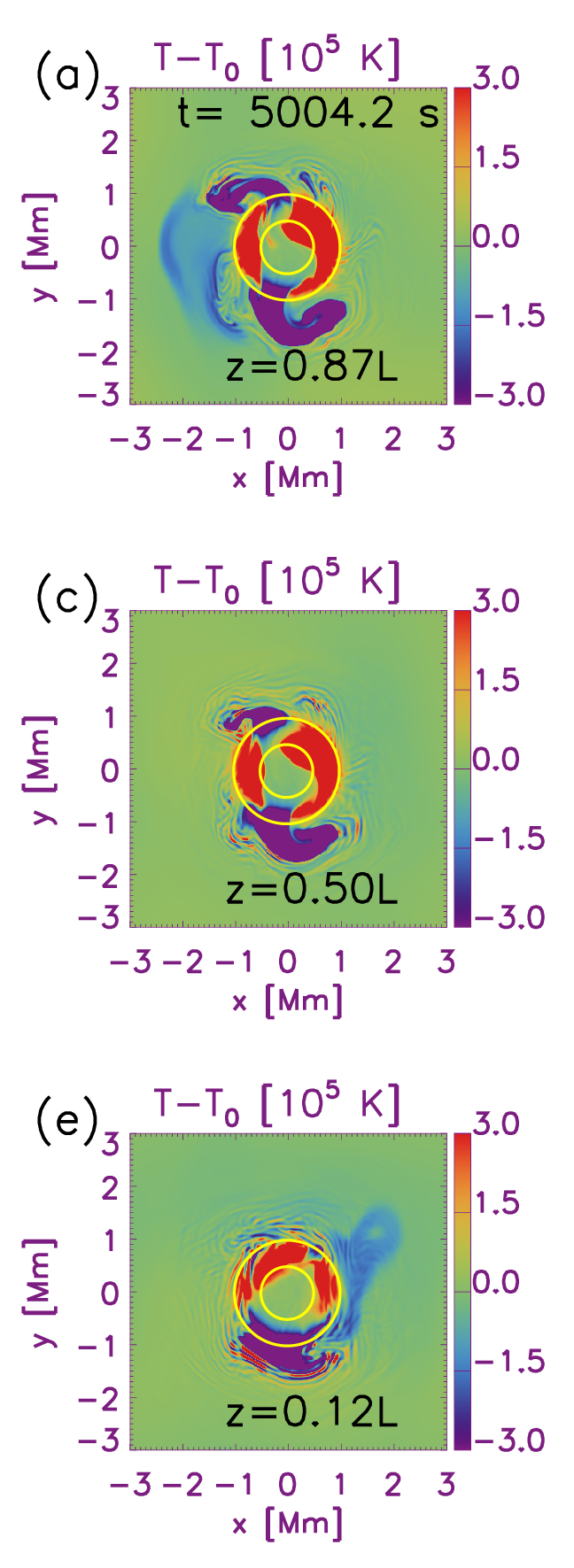} 
\includegraphics[scale=0.31]{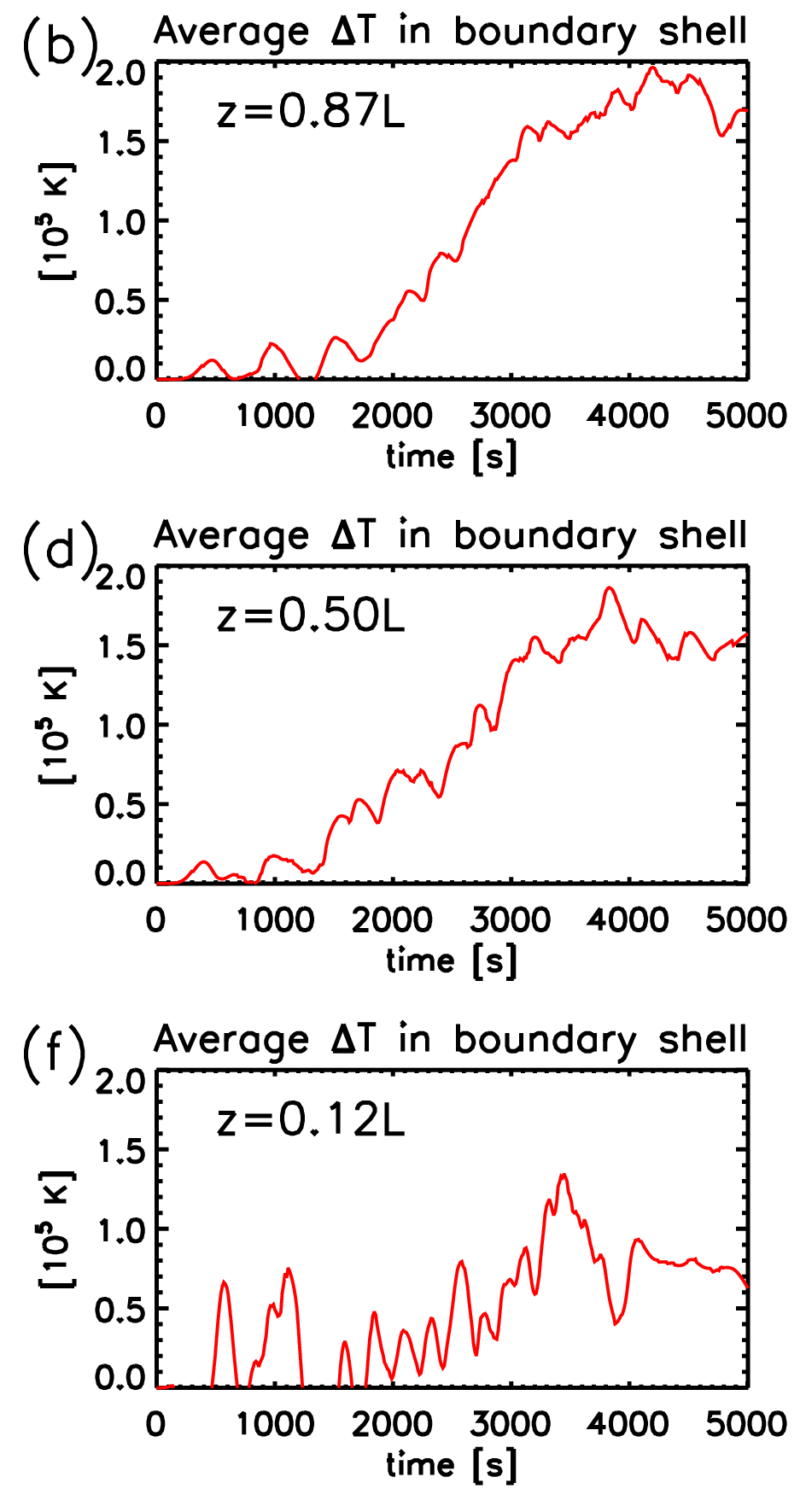} 
\caption{(a), (c), (e) are maps of the temperature variation over the simulation
at three cross sections near footpoint, at centre and near the end of the loop.
(b), (d), (f) show the time evolution of the
temperature variation on the boundary shell averaged over the density at the cross sections shown in the left hand side column.}
\label{tempzstime}
\end{figure}

Fig.\ref{tempzstime} shows the temperature increase in cross sections at the footpoint (Fig.\ref{tempzstime}e), halfway along the loop (Fig.\ref{tempzstime}c) and near the top boundary (Fig.\ref{tempzstime}a).
We highlight the boundary shell around the local centre of mass of the loop and we plot in
Fig.\ref{tempzstime}b, Fig.\ref{tempzstime}d, Fig.\ref{tempzstime}f
the average temperature increase in the boundary shell for each of these locations.
We find that near the footpoint (Fig.\ref{tempzstime}f), the temperature increase is very modest as it is where perturbations are generated and do not have sufficient time to dissipate, whereas
the temperature increase becomes more visible for the other two locations
(Fig.\ref{tempzstime}b, Fig.\ref{tempzstime}d),
where the temperature starts increasing and then it saturates when
it is about $1.5-2\times10^{5}$ $K$ above the initial temperature.
We also find that while around the cylinder there are both colder and hotter regions, the boundary shell largely presents regions with increased temperature. 

Furthermore, in order to isolate the thermal energy deposited by the dissipation of currents generated in the system (whether following phase-mixing or the velocity shear) we compare two simulations, the one here presented where we have $\eta=10^{10}\eta_S$,
and an additional run that differs from this one only for using the ideal MHD equation with $\eta=0$.
By comparing these two simulations we rule out the influence of numerical dissipation on our results and we also isolate the role 
of proper transverse wave energy dissipation by focusing on the kinetic energy associated with transverse velocities.
\begin{figure}
\centering

\includegraphics[scale=0.23]{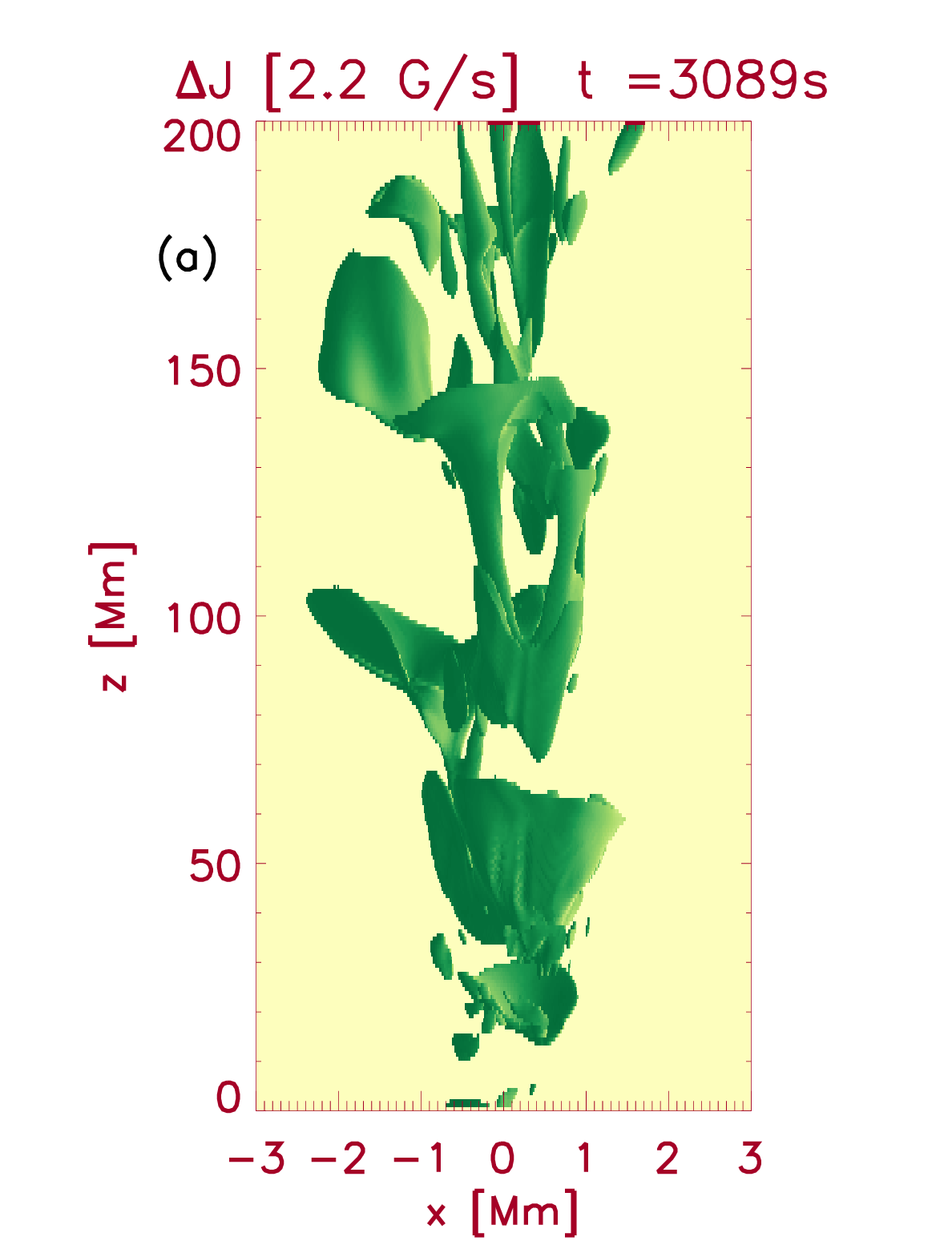} 
\includegraphics[scale=0.23]{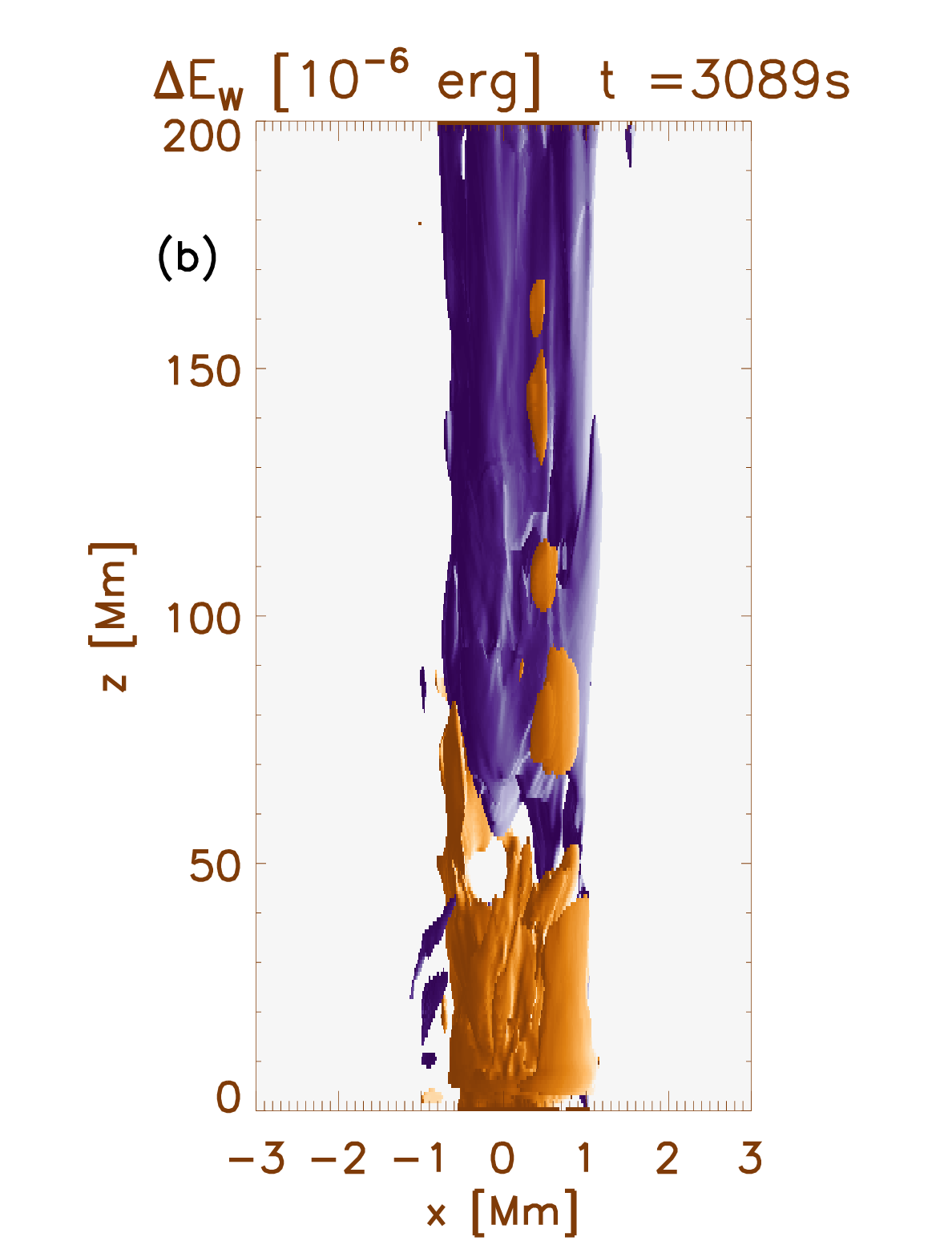} 
\caption{3D contours of the differences between the simulations with  $\eta=0$ and $\eta=10^{10}\eta_S$.
(a) is the electric current difference that shows where the simulation without resistivity has stronger currents. The differences contour is $10 j_0$.
(b) is the wave energy difference, where violet contours shows where the wave energy has been dissipated
in the simulation with resistivity and yellow contours shows where it occurs that the simulation with resistivity presents more wave energy.
The temporal evolution is available in a movie online.}
\label{compjew}
\end{figure}
We define the transverse wave energy as the kinetic and magnetic energy associated with the transverse components of the velocity and magnetic field.
Thus:
\begin{equation}
E_W=\frac{1}{2}\left(\rho v_x^2+\rho v_y^2 +\frac{B_x^2}{8\pi}+\frac{B_y^2}{8\pi}\right)
\end{equation}
The wave energy is generated in the domain from the transverse motion near the footpoint (lower boundary) and it decreases when the transverse perturbations are dissipated or leave the domain through the upper boundary.
Fig.\ref{compjew}a shows the 3D contours of the electric current difference between the two simulations at $t=3089$ s, which is a representative snapshot of the simulation. We find that the current distribution is different in the two simulations and that differences are mostly localised around the boundary shell and at higher z-coordinates. Near the footpoint the differences are less evident because currents have not been dissipated yet in the simulation with resistivity.
Correspondingly, in Fig.\ref{compjew}b we show the 3D contour of the difference in wave energy between the two simulations, where we use two colours
to indicate negative (yellow, more wave energy in the simulation with resistivity) or positive (violet, more wave energy in the simulation without resistivity) energy difference.
We find that at lower $z$ coordinates, positive and negative regions alternate (see movie) as the wave energy is just distributed differently on the $x-y$ planes,
whereas at higher $z$, the simulation without resistivity consistently shows more wave energy, 
because this is more efficiently dissipated in the simulation with resistivity.

\begin{figure}
\centering
\includegraphics[scale=0.28]{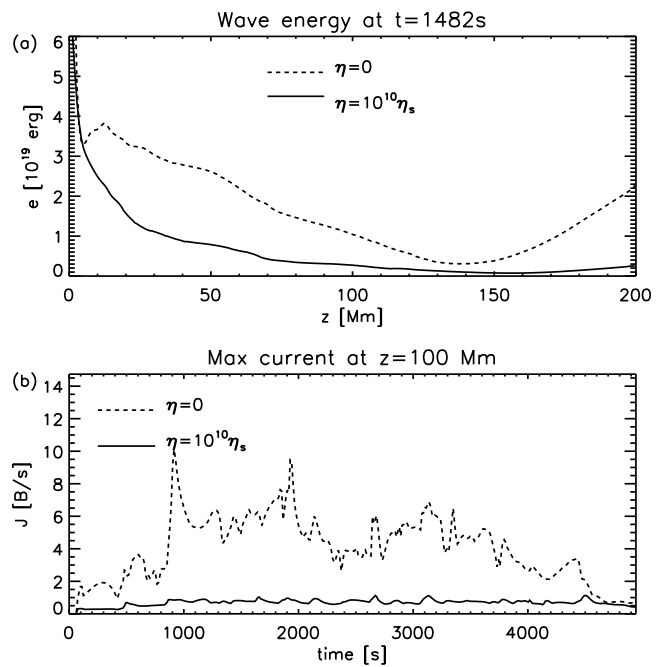} 
\caption{(a) wave energy as a function of z at $t=1482$ s for the simulation without resistivity and with resistivity.
(b) maximum of the current modulus at $z=100$ Mm (middle of the loop) as a function of time
for the simulations with and without resistivity.}
\label{reswavej}
\end{figure}

Fig.\ref{reswavej}a shows the average wave energy in the boundary shell at a single time ($t=1482$ s) for both 
simulations with and without resistivity, where we clearly find that while the simulation with $\eta=0$ allows the wave energy to increase along the domain 
due to the passage of different pulses, when resistivity is active the wave energy monotonically decreases along the loop.
As explained, this is connected with the dissipation of electric currents and Fig.\ref{reswavej}b shows the maximum current in the domain as a 
function of time, where we find that a resistivity of $\eta=10^{10}\eta_S$ is effective enough to quench all the currents at the threshold level to trigger anomalous resistivity, whereas when resistivity is not active stronger currents are allowed to build.

\begin{figure}
\centering
\includegraphics[scale=0.28]{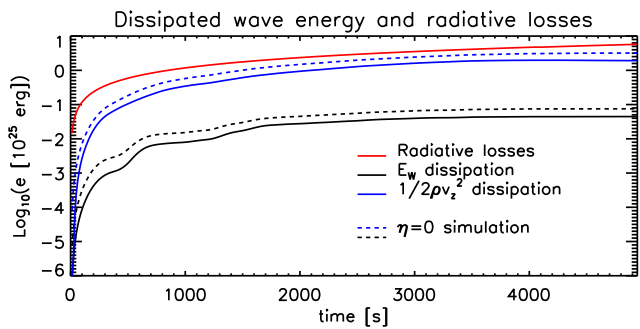} 

\caption{Continuous lines showing the energy dissipated in the boundary shell (as transverse waves energy and slow-modes energy) as a function of time compared with the radiative losses.
Dashed lines show the total energies in the simulation with $\eta=0$.}
\label{ethesolspe}
\end{figure}

In order to quantify to what extent this process is relevant for the coronal heating problem, we need to compare these energetic considerations
with the radiative losses that the coronal plasma would be subject to over the same time span.
Fig.\ref{ethesolspe} shows the integrated wave energy dissipation in the boundary shell as a function of time, against the integrated radiative losses in the same domain.
The radiative losses are estimated from the density and temperature of the boundary shell and the estimation does not significantly change when we vary the value of density and temperature within the range of the values found in the boundary shell.
In these circumstances, the actual dissipation of wave energy remains about 2 orders of magnitude smaller than the energy requirements from radiative losses. While this result is certainly affected by the input, we also find that to dissipate all the wave energy in the solar corona is not an obvious process,
as in our study, despite of the strongly enhanced magnetic resistivity, only 60\% of the wave energy is dissipated within the timeframe of the simulation.
For completeness, we also include the dissipation of the kinetic energy associated with the parallel perturbations, i.e. $\frac{1}{2}\rho v_z^2$.
This energy is comparable with the radiative losses, but still not enough to account for the energy budget of the boundary layer by a factor of $\sim3$.
Dashed lines in Fig.\ref{ethesolspe} show the evolution of wave energy and kinetic energy associated with perturbations in the simulation without resistivity.
As this is the maximum amount of energy that we can dissipate and as this does not match the estimation of the radiative losses,  
we find that in our simulation the footpoint motions are not sufficiently strong to supply enough energy to maintain the thermal structure of the loop. As both velocities and densities considered in this model are realistic for the solar corona, this seems a legitimate concern.

\subsection{Loop structure}
\label{loopstructure}

In addition to the impact of the propagating transverse waves on the thermal properties of the loop,
it is useful to investigate the effect on the loop structure itself. 
It is known that idealised Alfv\'en waves are incompressible, however it is reasonable to expect that
small non linear terms integrated over a sufficiently long time
and the interference between single pulses, that are not ideal Alfv\'en waves, play a role in modifying the loop structure. 

\begin{figure}
\centering

\includegraphics[scale=0.26]{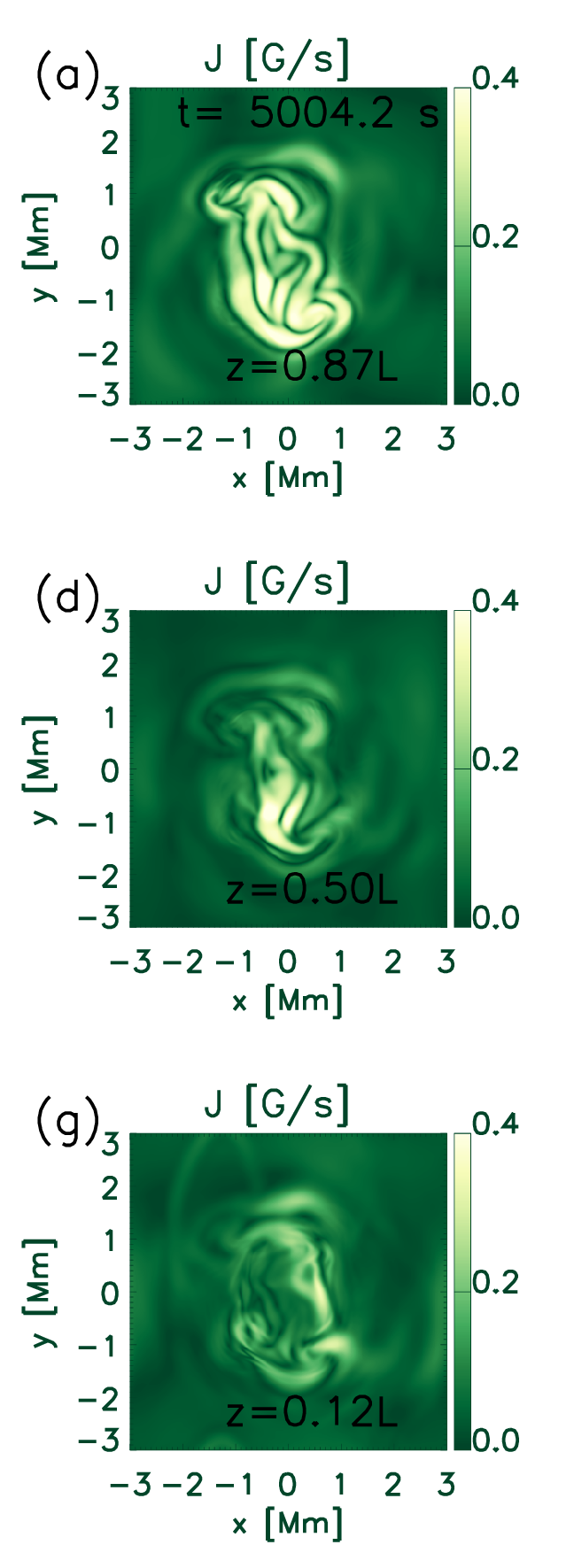} 
\includegraphics[scale=0.26]{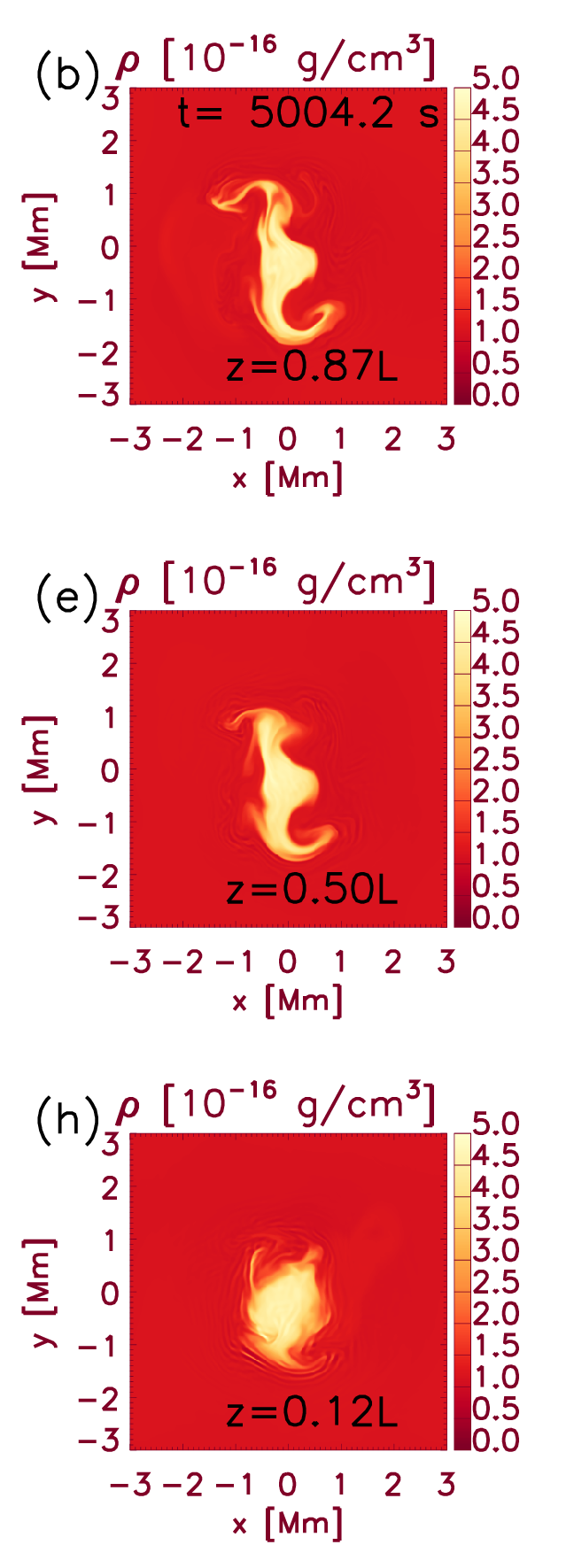} 
\includegraphics[scale=0.26]{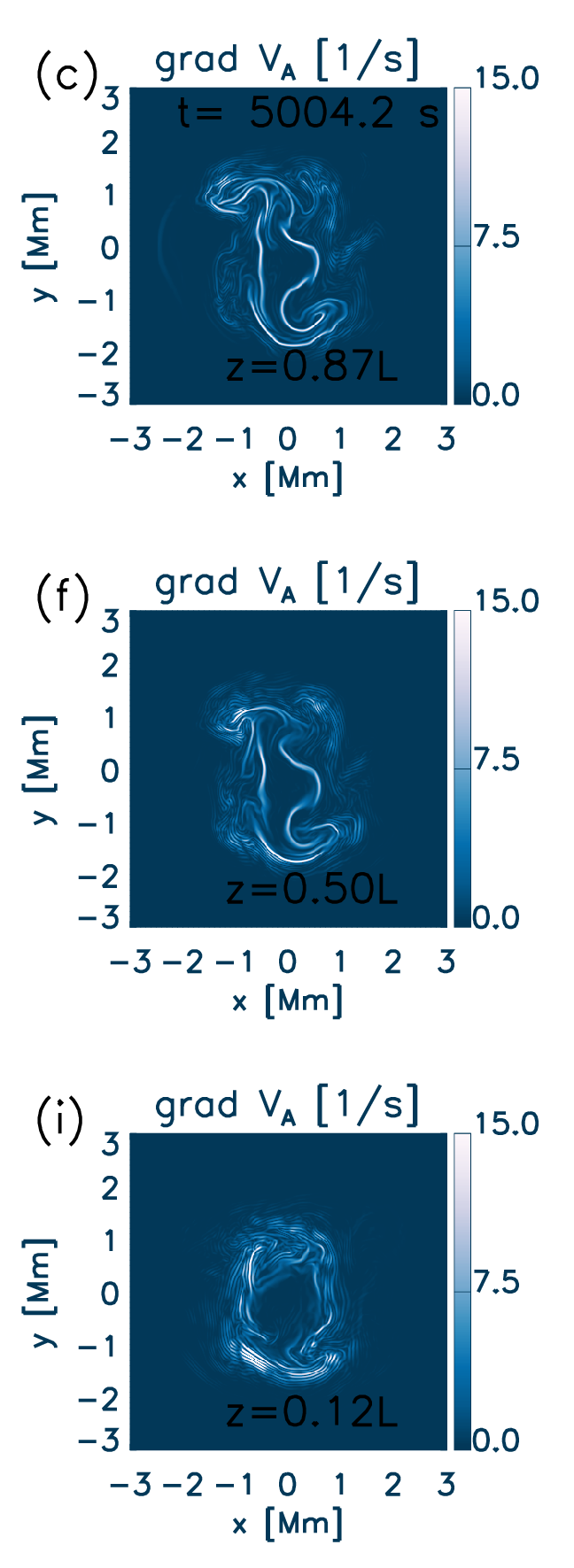} 

\caption{Cross sections of electric current modulus (a, d, g),  density (b, e, h), and module of 
the Alfv\'en speed gradient (c, f, i) near the foot points (g, h, i), at the middle of the loop (d, e, f),
and near the end the loop (a, b, c).
The temporal evolution is available in a movie online.}
\label{horizrhoj}
\end{figure}

KHI has been observed and modelled in coronal loop over the past few years by, for example, \citet{Terradas2008,Antolin2015,Karampelas2017,Howson2017}. However because of the limitation of current instruments, the discussion is not settled yet whether this is a common phenomenon in coronal loops or rather events limited to more energetic oscillations.
Our simulation is not a favourable configuration to trigger KHI for a number of reasons.
Firstly, we have a rather modest wave energy input from one footpoint and secondly the other domain boundary is not linetied,
thus we do not have reflection of the waves, and standing modes do not settle in.
KHI is more likely to occur in the presence of consistent oscillations of the cylinder and thus higher shear velocities between the cylinder and the background medium.
Thirdly, the magnetic resistivity is known to quench the development of KHI \citep{Howson2017}.
Nevertheless, KHI develops as soon as $\sim 1000$ s near the footpoint and slightly later higher up in the loop.
Fig.\ref{horizrhoj} shows the electric current and density distribution at different cross sections of the domain at the final time of the simulation.
We find that both currents and density reach a highly structured distribution that naturally leads to a very heterogeneous loop structure. 
It is remarkable that the idealised cylindrical structure has been significantly affected by the persistent propagation of waves and 
in the end the loop structure is hardly recognisable. 
This also addresses the matter of the life span of coronal loops, where our simulation shows that after a time span of the order of hours
the loop structure becomes significantly altered and might completely disappear,
even without significant dissipation of energy.

Moreover, Fig.\ref{horizrhoj}c, Fig.\ref{horizrhoj}f, Fig.\ref{horizrhoj}i shows the final gradient of the Alfv\'en speed in the transverse direction at the same cross sections.
We find that the development of KHI has significantly affected the distribution of the gradient of the Alfv\'en speed, leading
to a distortion of the region where phase-mixing can occur.
Initially, the region where the gradient of the Alfv\'en speed is significant is a ring around the central axis of the loop (the boundary shell),
where the intensity of the gradient of the Alfv\'en speed is $1$ s$^{-1}$.
At the end, the ring structure is lost and the gradient of the Alfv\'en speed is steeper at the edge of the KHI structures and it has gone up to $20$ s$^{-1}$.
However, the rate at which the wave energy is converted does not seem to undergo any significant change during the simulation, as while the gradient of the Alfv\'en speed becomes steeper, the region where the Alfv\'en speed varies shrinks.

\subsection{Two footpoints driver simulation}
\label{twofootpoints}

To complete our modelling of how the propagation of transverse waves induced
by the footpoint displacement in coronal loops affects the thermal structure of the loop,
we have also investigated the configuration of a magnetised cylinder (ignoring curvature) where both footpoints are attached to the base of the solar corona and driven simultaneously.
With this regard, we run an MHD simulation that is the same as the one described thus far with the exception that the upper boundary is now also driven by a displacement and velocity profile that although different, is constructed in the same way as the driving profile used on the lower boundary. Hence, in this simulation, the waves generated at each footpoints do not leave the domain at the opposite boundary, but interact with the waves incoming from the other side. Apart from the fact that more energy enters the domain, this scenario allows us to investigate whether the interaction of waves travelling in opposite directions leads to more efficient energy dissipation.

\begin{figure}
\centering
\includegraphics[scale=0.23]{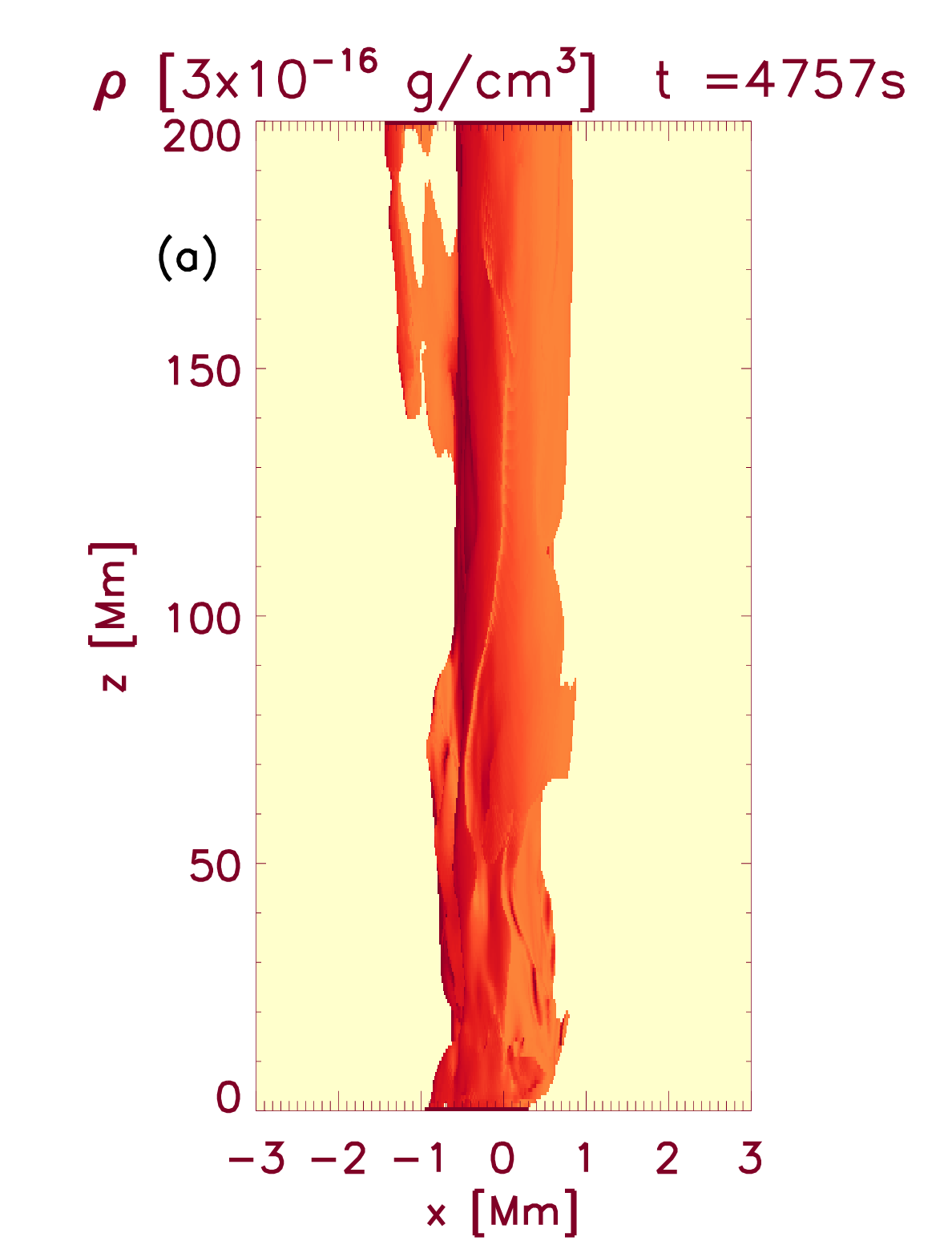}
\includegraphics[scale=0.23]{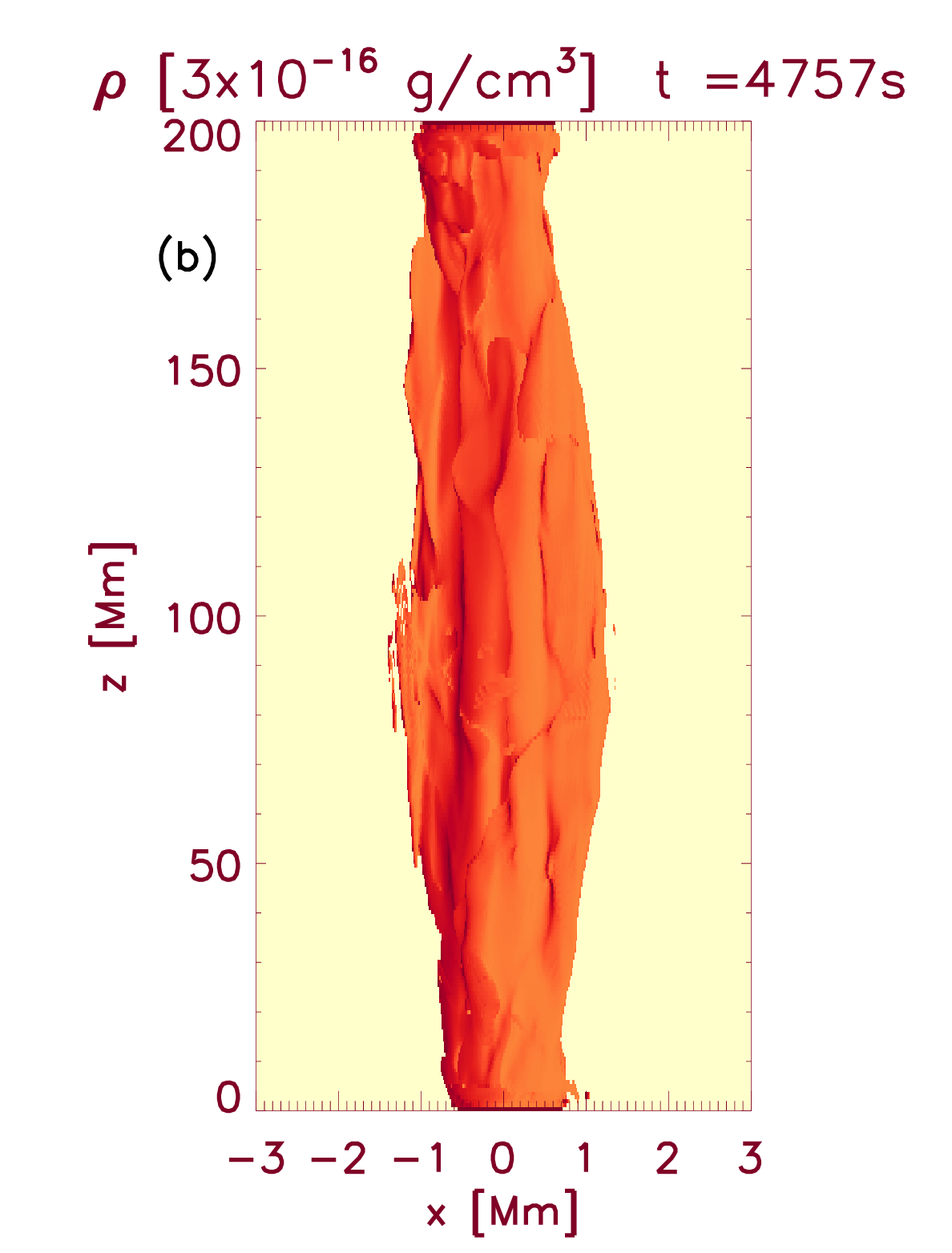}

\caption{3D contours of density at $t=4757$ s for the simulation
where we drive only one footpoint (a) and where we drive both footpoints (b).
The temporal evolution is available in a movie online.}
\label{solspe2ftdens}
\end{figure}

Fig.\ref{solspe2ftdens} compares the 3D contour of density of 
the simulation with two footpoints driven (Fig.\ref{solspe2ftdens}b) with the simulation
in which only one footpoint is driven (Fig.\ref{solspe2ftdens}a) at $t=4757$ s.
We find that in both cases, the density structure has significantly evolved,
but while structures have visibly different sizes
in the simulation where only one footpoint is driven,
the density distribution appears more symmetric when both footpoints are driven and 
the middle of the cylinder is significantly more expanded than the footpoints due to
the development of the KHI instabilities.
However, the main features are similar in both simulations and
the interaction of counter-propagating waves does not significantly 
alter the extension of the structures that are formed through the continuous passage of waves along the loop.

In order to verify if the interaction of counter propagating waves leads
to higher energy deposition,
we plot the wave energy dissipation for both simulations as a function of time.
As explained in Sect.\ref{solarsimulation} this is computed from the comparison between
a simulation with $\eta=0$ and one with a finite value for $\eta$. 
We do not run a simulation with $\eta=0$ when two footpoints are driven,
so in order to estimate the wave energy in this simulation we just multiply 
the wave energy in the simulation with $\eta=0$ and one driven footpoint  
by a factor $\zeta$ that is computed from the wave energy integral of the drivers.
\begin{equation}
\zeta\left(t\right)=\frac{\int^{t}_{0}  u_1^2\left(t \right) + u_2^2\left(t \right) \, dt}{\int^{t}_{0}  u_1^2\left(t \right)  \, dt}
\end{equation}
where $u_1\left(t \right)$ and $u_2\left(t \right)$ are 
the velocity amplitude of the two drivers.

\begin{figure}
\centering
\includegraphics[scale=0.28]{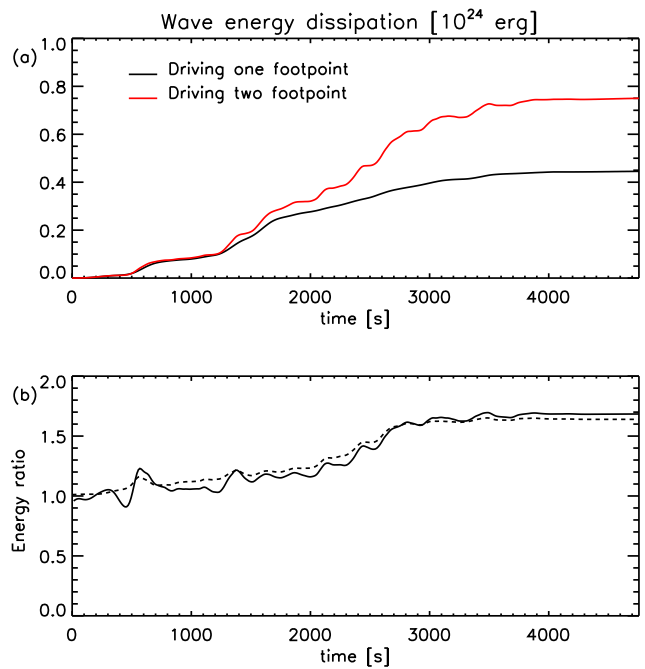}

\caption{(a) Wave energy dissipation in the two simulations where we drive either one or two footpoints. 
(b) Ratio between the energy dissipation in the two simulations (continuous line)
and between the energy of driver in the two simulations (dashed line).}
\label{ethesolspefoot}
\end{figure}

Fig.\ref{ethesolspefoot}a shows the dissipation of the wave energy as a function of time.
We find that the simulation where both footpoints are driven certainly dissipates more energy
and that the two simulations diverge in time until the difference reaches a constant value 
when the drivers stop.
However, when we plot in Fig.\ref{ethesolspefoot}b the 
ratio between the energy input in the two simulations
against the ratio between the energy dissipated in the two simulations
we find that these two quantities are highly correlated.
This shows that, at least in this regime and with this numerical resolution, the interaction between counter-propagating waves 
does not lead to a more efficient dissipation of the wave energy, 
as the difference between the two simulation is only due to the different energy input.

\section{Discussion and Conclusions}
\label{discussions}

In this paper, we have analysed how the propagation of transverse MHD waves along a coronal loop can affect the structure and the thermodynamics of the loop.
Previous studies have addressed the energy deposition in the solar corona due to the propagation of Alfv\'en waves, especially in the framework of phase-mixing, and have concluded that in the presence of Alfv\'en speed gradients some of the wave energy is converted into heating
\citep{Pascoe2010,Pascoe2011,Pascoe2012,PaganoDeMoortel2017}.
A key feature of our investigation is to use a model for the transverse motions spectrum observed in the solar corona by \cite{Morton2016}.
We have selected the spectrum and the modelling parameters to best represent active regions loops.
By using such a driver we aim to simulate the effect of the buffeting of the coronal loop footpoints
on the propagation of waves and the structure of the loop itself. Although using an observed power spectrum allows us to draw some conclusions on the effectiveness of this process in heating the solar corona, our idealised loop structure and the dynamic model still depart from a proper description of a coronal loop system.
For instance, as an example, \citet{LopezAriste2015} explained that a minor contribution of sausage modes should be accounted for.

We have initially built a model that is motivated by an observed spectrum of transverse oscillations in the solar corona by means of superposition of a number of single pulses with different duration, amplitudes and with a random direction
to generated a random walk of the loop footpoint.
As the observed power spectrum decreases with frequency, we find that it can be reproduced when the long period pulses (longer than some minutes) oscillate
at a speed about one order magnitude larger than the high-frequency ones. 
The spectrum used in this work is measured from a direct long observation of the solar corona, and it accounts for both energy input to the power spectrum
\citep[probably from the chromosphere, e.g.][]{2002ApJ...564..508R,2003ApJ...599..626B,2008SoPh..251..251C,2012ApJ...746...68K,Jess2009,2015ApJ...814..106C}
as well as energy output
\citep[from the damping of waves or their leaking to higher layers of the solar atmosphere, e.g.][]
{Morton2014,Goddard2016}.
In other words, this can be considered as a steady state spectrum which, if Alfv\'en waves were directly responsible for the coronal heating, results from the balance of energy dissipated to maintain the million degree corona and (wave) energy supplied to the corona. If this steady-state power spectrum fails to supply a sufficient amount of energy to the solar corona, it means that a different mechanism is needed to provide the additional energy.

Alfv\'en waves need to propagate over a distance significantly larger than their wavelength before phase-mixing develops the required transverse small scales where the dissipation of the wave energy can occur efficiently.
Therefore, we investigate first the dynamics of the propagation of transverse waves of different wavelengths selected from our representative set of pulses to understand how different frequencies can lead to the deposition of thermal energy due to the dissipation of electric currents.
We find that the currents generated by the propagation of transverse waves are crucially dependent on the frequency of the pulse. 
While short-period oscillations can induce significant electric currents through phase-mixing, this is not the case for long-period oscillations whose wavelength is comparable with the loop length.
However, we find that for our input power spectrum, the higher velocity amplitudes of the long-period oscillations lead to comparable currents generated in slow-mode waves following the compression of the guide magnetic field.
For this to occur, it is sufficient to have a relative motion between the loop and the background corona, which is a likely scenario as velocity fields in the photosphere and chromosphere are usually of the same spatial scale as the loop (flux tube). Such coexistence of propagating fast transverse waves and slow modes was observed by e.g.~\cite{Threlfall2013}.
The currents generated in either way have similar strength,
but are found at different loop locations.
The ones associated with phase-mixing are generated away from the footpoint,
while the ones associated with slow-modes are instead formed very close to the footpoints.
Additionally, we find that in our description single pulses remain in a linear regime and
do not carry enough energy to lead to noticeable heating.

Because of the compromises we need to make between the long temporal duration of the MHD numerical simulation 
and the spatial resolution of the numerical model we adopt a technique where we use an anomalous resistivity to efficiently dissipate the wave energy once the phase-mixing starts.
We find that such an approach is a good compromise for our purposes and it preserves 
the key features of the actual resistive processes in the solar corona.
The resistivity used in this model is $10^{10}$ times the resistivity prescribed by \citet{Spitzer1962} that describes the diffusion of 
the magnetic field in the MHD regime. However, the spatial scales considered in this work are significantly larger than the smallest spatial scale 
where the MHD approximation still holds. This means that we are probably underestimating the magnetic field gradients.
Let us assume that our MHD model properly estimates the induction equation diffusion term,
then for one given component of the magnetic field (e.g. $B_i$) and in one direction (e.g. $x$) this would mean that:
\begin{equation}
\eta_S\frac{c^2}{4\pi}\frac{\partial^2 B_i}{\partial x^2}\bigg\rvert_{MHD}\sim\eta\frac{c^2}{4\pi}\frac{\partial^2 B_i}{\partial x^2}\bigg\rvert_{Sim}
\end{equation}
where the subscripts MHD and Sim refer to the actual gradients in the solar corona and in the MHD simulation, respectively.
For the two terms to be comparable, it means that the Laplacian of the magnetic field must compensate for the different resistivity coefficients.
This can be justified if $\Delta x\rvert_{MHD}=10^{-5}\Delta x\rvert_{Sim}$.
As in our simulations $\Delta x\rvert_{Sim}=15$ km, the smallest MHD scales where the dissipation of currents happen is $\sim15$ $cm$.
This argument leads to reasonable spatial scales and justifies using such an enhanced resistivity in this model.

Our 3D MHD simulation of the coronal loop when we use a combined driver that consists of the pulses to generate the observed power spectrum suggests that the conversion of wave energy into heating cannot satisfy the energy requirements to maintain a million degrees solar corona.
In order to come to these conclusions we have run two simulations with and without resistivity to identify the thermal energy exchange that is due to the dissipation of transverse wave energy. We find that this accounts only for about $1\%$ of the radiative losses of the boundary shell of the loop.
It slightly increases if we account also for the dissipation of the longitudinal motions generated by the slow-mode waves.
Moreover, it should be noted that our MHD model shows transverse velocities of the order of $1$ km/s
in the solar corona. Such amplitudes, even if generated from a driver that shows
considerably higher velocities, are lower than what has been measured by \citet{Threlfall2013} ($\sim10$ km/s),
but also lower than other estimates. For example,
\citet{McIntosh2011} report average values of $5$ km/s,
and \citet{MortonMcLaughlin2013} estimate values of about $3$ km/s.
In this scenario, one can conclude that the available kinetic energy to heat the plasma
can be one order of magnitude larger.
While this can make the wave heating contribution larger, this is not yet enough to balance the radiative losses.
Furthermore, \citet{Cargill2016} argue that the plasma would react to the phase-mixing heating
by shifting the density gradients and thus making the heating less efficient.
It remains possible that significant wave energy is deposited at lower layers of the solar atmosphere or in open field regions.
Alternatively, the results by \citet{LopezAristeFacchin2018} suggest that wave connected heating events can be localised in space and \citet{Reale2016} explain how the heating could be deposited in the lower atmosphere when loop structures are perturbed.
At the same time, it should be noted that the investigation we have carried out here is necessarily 
bound to a small region of the parameter space and more work is needed to extend this approach in regimes
representative of other, equally realistic, configurations that could favour a more efficient heating.
Parameters that we have not fully investigate here are
i) the density contrast between the loop and the background corona,
ii) the Alfv\'en speed that can range between $200$ $km/s$ to more than $1000$ $km/s$ \citep{McIntosh2011,2001A&A...372L..53N},
iii) power spectra derived from other regions, 
iv) the overall amplitude of the driver.

In terms of coronal heating models, we can summarise that
both short and long-period oscillations seem
to generate and dissipate currents.
While our magnetic anomalous resistivity is tuned to dissipate the wave energy (thus a relatively low threshold), future investigations could examine whether similar dynamics can be described in the framework of the braiding of coronal loops,
where multiple drivers act on the coronal loop footpoints entangling magnetic field lines.

Finally, a key results of this work is the development of KHI when the observed power spectrum is used to drive one footpoint of the loop.
While the initial density contrast in our simulation is rather high in order to enhance the phase-mixing of Alfv\'en waves, KHI develops quite quickly after the onset of the footpoint driver and does so even when only one footpoint is driven and in a rather diffusive regime.
All these elements indicate that the development of KHI could be a rather common phenomenon in coronal loops and
could influence the apparent (observable) life time of some coronal loops.
At the same time, the development of KHI leads to significantly higher gradients of the Alfv\'en speed but over a smaller volume.
The novelty of this result that adds to previous work
\citep[e.g.][]{BrowningPriest1984, Terradas2008, Antolin2015,2016A&A...595A..81M,Howson2017, Karampelas2017,PaganoDeMoortel2017,Pagano2018} is that KHI can be generated using the observed power spectrum without standing modes. This seems to indicate that a turbulent cascade could be triggered by Alfv\'en waves \citep{vanBallegooijen2011}, where
the development of KHI is just one step towards the creation of smaller scale structures
that starts with the propagation of Alfv\'en waves and enhances a faster dissipation of the waves themselves. Additionally, \citet{2017NatSR...714820M} argue that turbulence can develop also from propagating waves.
However, the limited spatial resolution of this simulation does not allow to access the full turbulent cascade that would require additional analysis. This is particularly true when we compare the simulations where both or one footpoints are driven, as we find here that the thermal energy deposition is linearly dependent on the energy input, that is a result that definitely requires more investigation.

To conclude, the present work expands the ongoing research into wave based heating mechanism(s) of the solar corona.
Its key feature is that it investigates the response of a coronal loop to an observed power spectrum of transverse oscillation in the solar corona, allowing us to study the effect of a realistic footpoint driver on the dynamics of a coronal loop.
At the same time, this approach helps remove some of the previous uncertainties of the wave based heating models and it indicates once more that the dissipation of transverse MHD waves via phase-mixing alone in the solar corona does not seem to be the answer to the coronal heating problem, at least in the investigated regime. However, the propagation of transverse of waves and their dissipation could be part of a more complex scenario that through braiding or turbulence could lead to the conversion of higher amounts of energy.

\begin{acknowledgements}
We would like to thank the referee for the constructive comments that have definitely contributed to improving the manuscript.
This research has received funding from the UK Science and Technology Facilities Council (Consolidated Grant ST/K000950/1) and the European Union Horizon 2020 research and innovation programme (grant agreement No. 647214). 
This work used the DiRAC Data Centric system at Durham University, operated by the Institute for Computational Cosmology on behalf of the STFC DiRAC HPC Facility. This equipment was funded by a BIS National E-infrastructure capital grant ST/K00042X/1, STFC capital grant ST/K00087X/1, DiRAC Operations grant ST/K003267/1 and Durham University. DiRAC is part of the National E-Infrastructure.
We acknowledge the use of the open source (gitorious.org/amrvac) MPI-AMRVAC software, relying on coding efforts from C. Xia, O. Porth, R. Keppens.
\end{acknowledgements}

\bibliographystyle{aa}
\bibliography{ref}

\end{document}